\documentclass[a4paper,fleqn,usenatbib,useAMS]{mnras}

\usepackage{graphicx}	
\usepackage{amsmath}	
\usepackage{amssymb}	
\usepackage{multicol}        
\usepackage{bm}		
\usepackage{pdflscape}	

\usepackage{multicol}
\usepackage{caption}
\usepackage{threeparttable}

\newcommand{\kms}{$\,$km$\,$s$^{-1}$}

\usepackage[T1]{fontenc}
\usepackage{ae,aecompl}

\title{High-Sensitivity Observations of Molecular Lines with the Arecibo Telescope}

\author[W. S. Tan et al.]{
W. S. Tan,$^{1}$
E. D. Araya,$^{1,*}$
L. E. Lee,$^{1, **}$
P. Hofner,$^{2,3}$
S. Kurtz,$^{4}$
H. Linz,$^{5}$
L. Olmi.$^{6}$
\\
$^{1}$ Physics Department, Western Illinois University, 1 University Circle, Macomb, IL 61455, USA.\\
$^{2}$ New Mexico Institute of Mining and Technology, Physics Department, 801 Leroy Place, Socorro, NM 87801, USA.\\ 
$^{3}$ Adjunct Astronomer at the National Radio Astronomy Observatory, 1003 Lopezville Road, Socorro, NM 87801, USA.\\
$^{4}$ Instituto de Radioastronom\'{i}a y Astrof\'{i}sica,
Universidad Nacional Aut\'{o}noma de M\'{e}xico, Apdo. Postal 3-72, 
58089, Morelia,\\ 
~ Michoac\'{a}n, Mexico.\\
$^{5}$Max Planck Institute for Astronomy, K\"onigstuhl 17, 69117, Heidelberg, Germany.\\
$^{6}$INAF, Osservatorio Astrofisico di Arcetri, Largo E. Fermi 5, I-50125 Firenze, Italy.\\
$^*$ Contact author: ed-araya@wiu.edu \\
$^{**}$ Currently at University of Missouri-Columbia.
}

\date{Accepted 2020 June 18. Received 2020 June 16; in original form 2019 September 12}

\pubyear{2020}

\begin{document}
\label{firstpage}
\pagerange{\pageref{firstpage}--\pageref{lastpage}}
\maketitle

\begin{abstract}
We report on one of the highest sensitivity surveys for molecular lines in the frequency range 6.0 to 7.4$\,$GHz conducted to date. The observations were done with the 305$\,$m Arecibo Telescope toward a sample of twelve intermediate/high-mass star forming regions. We searched for a large number of transitions of different molecules, including CH$_3$OH and OH. The low RMS noise of our data ($\sim 5\,$mJy for most sources and transitions) allowed detection of spectral features that have not been seen in previous lower sensitivity observations of the sources, such as detection of excited OH and 6.7$\,$GHz CH$_3$OH absorption. A review of 6.7$\,$GHz CH$_3$OH detections indicates an association between absorption and radio continuum sources in high-mass star forming regions, although selection biases in targeted projects and low sensitivity of blind surveys imply incompleteness. Absorption of excited OH transitions was also detected toward three sources. In particular, we confirm a broad 6.035$\,$GHz OH absorption feature in G34.26+0.15 characterized by an asymmetric blue-shifted wing indicative of expansion, perhaps a large scale outflow in this H{\small II} region.  
\end{abstract}





\section{Introduction} \label{Introduction}

In the past two decades, high sensitivity receivers have been deployed in major observatories around the world for observations at 6$\,$GHz frequencies. Different molecular transitions have been detected at these frequencies, particularly in regions of high-mass star formation. The main examples are excited OH transitions (e.g., \citealt{Avison_2016MNRAS.461..136A}, \citealt{Al-Marzouk_2012ApJ...750..170A}) and the widespread 6.7$\,$GHz methanol line. 

The 6.7$\,$GHz methanol transition was first detected in the interstellar medium by \cite{1991ApJ...380L..75M} using the 140 foot telescope of the NRAO. \cite{1991ApJ...380L..75M} detected strong methanol masers associated with star forming regions, as well as methanol absorption toward eight regions (NGC$\,$2264, G345.70$-$0.09, NGC$\,$6334$-$C, Sgr A$-$F, Sgr A$-$A, Sgr B2, G10.62$-$0.38, W33). Since then, most observations have focused on studies of the 6.7$\,$GHz CH$_3$OH maser as part of blind surveys (e.g., \citealt{Breen_2015MNRAS.450.4109B}; \citealt{2007ApJ...656..255P}), targeted observations of high-mass star forming regions (e.g., \citealt{Olmi_2014AA...566A..18O}), or to investigate periodic maser flares (e.g., \citealt{Rajabi_2019MNRAS.484.1590R}, \citealt{MacLeod_2018MNRAS.478.1077M}, \citealt{2018MNRAS.474..219S}, \citealt{araya10}, \citealt{2004MNRAS.355..553G}). While most observations reported in the literature have focused on bright lines, high-sensitivity observations have resulted in detection of weak lines, such as 6.7$\,$GHz methanol absorption in the low-mass star forming region NGC$\,$1333 \citep{Pandian_2008AA...489.1175P}. In this case, the line is due to an anti-inversion (over-cooling) effect that allows absorption against the Cosmic Microwave Background (CMB), as observed for example in 6$\,$cm H$_2$CO (e.g., \citealt{Araya_2006AJ....132.1851A}) and 12$\,$GHz CH$_3$OH transitions (\citealt{Walmsley_1988AA...197..271W}, \citealt{Peng_1991PASAu...9..287P}). \cite{Pandian_2008AA...489.1175P} concluded that the methanol absorption in NGC$\,$1333 is tracing dense ($\sim 10^6\,$cm$^{-3}$) cold/warm ($\sim 15$ to $30\,$K) gas. 

In extragalactic environments, high-sensitivity observations have shown that 6.7$\,$GHz methanol absorption is rare~\citep{Ellingsen_1994MNRAS.267..510E}, e.g., it was not detected in the Karl G. Jansky Very Large Array (VLA) observations of M31 by \cite{Sjouwerman_2010ApJ...724L.158S}. Extragalactic 6.7$\,$GHz methanol absorption has only been reported towards the nucleus of the active galaxy NGC$\,$3079 \citep{2008AA...484L..43I}, the ultra-luminous infrared galaxy (ULIRG) Arp 220 \citep{Salter_2008AJ....136..389S}, and a tentative detection toward Mrk$\,$348 \citep{Impellizzeri_2008_PhD}. 

Previous high-sensitivity and high-spectral resolution observations of CH$_3$OH and other molecular species have shown the potential of high-sensitivity studies for detection of weak variable masers, narrow lines tracing quiescent molecular clouds, and absorption features with high-velocity linewidths that could trace outflows (e.g., \citealt{Al-Marzouk_2012ApJ...750..170A}, \citealt{Araya_2006AJ....132.1851A}). Indeed, a great analytic potential is open when lines are found in absorption against continuum provided by target sources. Combined with high-sensitivity and good velocity resolution, the kinematics of the gas along the line-of-sight to the targets can be probed, and the absorption nature removes a geometric ambiguity regarding the interpretation. 

Motivated by the potential of high-sensitivity and high-spectral resolution studies, we conducted observations of a sample of twelve intermediate and high-mass star forming regions with the 305$\,$m Arecibo Telescope in Puerto Rico. Our observations were carried out in two main runs as described in section \ref{Observations and Data Reduction}. Results and discussion are presented in sections \ref{Results} and \ref{Discussion}, respectively, where we focus on 6.7$\,$GHz CH$_3$OH absorption and excited OH lines. Our conclusions are given in section \ref{Summary}.\footnote{This work is based on the MS Thesis ``Survey for C-Band High Spectral Lines with the Arecibo Telescope'', by \cite{Tan2017}.}

\section{Observations and Data Reduction} \label{Observations and Data Reduction}

\subsection{Main Survey} \label{mainsurvey}

The 305$\,$m Arecibo Telescope in Puerto Rico was used to observe twelve intermediate and high-mass star forming regions in the first Galactic quadrant (Table~\ref{tb_sources}). The sources were mostly selected from two main samples: high-mass protostellar candidates from \cite{Sridharan_2002ApJ...566..931S}, which were observed by \cite{Araya_2004ApJS..154..579A}, and regions known to harbour ultra-compact (UC) H{\small II} regions and prominent hot molecular cores (\citealt{Wood_1989ApJS...69..831W}, \citealt{Kurtz_2000prpl.conf..299K}), i.e., regions similar to G31.41+0.31 (e.g., \citealt{Araya_2008ApJ...675..420A}). Specifically, six sources (IRAS$\,$18472$-$0022, G34.26+0.15, G45.07+0.13, G45.12+0.13, G45.47+0.05, G69.54$-$0.98) have been classified as UCH{\small II} regions (\citealt{Wood_1989ApJS...69..831W}; \citealt{Kurtz_1994ApJS...91..659K}; \citealt{Hoare_2012PASP..124..939H}, \citealt{Purcell_2013ApJS..205....1P}), four sources (IRAS$\,$18517+0437, IRAS$\,$18566+0408, IRAS$\,$19012+0536, IRAS$\,$19266+1745) are characterized by weak radio continuum emission at few mJy to sub-mJy levels \citep{Rosero_2016ApJS..227...25R}, and two sources (IRAS$\,$20051+3435, IRAS$\,$20081+2720) have 3.6$\,$cm flux density upper limits of 100$\,$mJy from \cite{Sridharan_2002ApJ...566..931S} and may be intermediate-mass star forming regions based on their FIR luminosity ($2\times 10^2$ or $10^4\,$L$_\odot$ for IRAS$\,$20051+3435 depending on kinematic distance; $3\times 10^2\,$L$_\odot$ for IRAS$\,$20081+2720; \citealt{Sridharan_2002ApJ...566..931S}). No homogeneous mass determination is available for all sources in our sample, however, gas masses (estimated from dust continuum and/or molecular gas observations; \citealt{Miralles_1994ApJS...92..173M}, \citealt{Beuther_2002ApJ...566..945B}, \citealt{Williams_2004A&A...417..115W}, \citealt{Cesaroni_2015A&A...579A..71C}) are above $\sim$600$\,$M$_{\odot}$ for all but two sources: IRAS$\,$20051+3435 (95 or 510$\,$M$_\odot$ depending on kinematic distance; \citealt{Beuther_2002ApJ...566..945B}) and IRAS$\,$20081+2720 ($\sim$$10\,$M$_\odot$, \citealt{Beuther_2002ApJ...566..945B}). 

We used the C-Band High receiver and the Wideband Arecibo Pulsar Processor (WAPP) spectrometer to observe 13 molecular transitions and one radio recombination line in two different configurations, each one with its own set of spectral windows. Configuration 1 was observed on October 13 and 14, 2008. Configuration 2 was observed on October 27, 2008. Table~\ref{tb_config} lists the specific spectral lines (grouped by molecule), upper energy level of the transitions, final channel width after smoothing (smaller channel widths were used in the case of narrow lines, including tentative detections), whether a spectral line was detected, and the configuration. All spectral windows were observed with a bandwidth of 3.125$\,$MHz (corresponding to velocity ranges between 125 to 155\kms~at the observed frequencies), 9 level sampling, dual linear polarization, and 2048 channels (1.5$\,$kHz initial channel spacing). The sources were observed in ON-OFF (position-switched mode) with 5 minutes integration ON-source per scan; two scans were obtained per source in most cases. All data reduction was done in IDL\footnote{IDL (Interactive Data Language) is a trademark of Harris Geospatial Corp.} using the Arecibo Observatory (AOIDL) routines.

The position-switched observations allowed us to correct the ON-source spectra for bandpass structures and to measure the radio continuum from the bandpass level (e.g., see \citealt{Watson_2003ApJ...587..714W}), which is reliable as long as the reference position samples a similar extended background radiation and has no compact bright continuum source. In our experience, radio continuum measurements with the Arecibo Telescope from position-switched observations are reliable above $\sim 100\,$mJy; in lines-of-sight with low confusion levels, reproducible continuum measurements below $100\,$mJy have been obtained (e.g., \citealt{Strack_2019ApJ...878...90S}).

\begin{table*}
\begin{center}
\begin{threeparttable} 
\centering
\caption{Source List}
\label{tb_sources}
\begin{tabular}{lcc ccc} 
\hline
Source           & l        & b	  & R.A. (J2000) & Decl. (J2000)             & Config. \\
                 & ($^{\circ}$) & ($^{\circ}$)& (h m s)& (\degr~ \arcmin~ \arcsec) &     \\
\hline
IRAS$\,$18472$-$0022$^a$ & 32.466   & 0.209   & 18 49 50.70  & $-$00 19 09.0 & 1   \\
G34.26+0.15              & 34.257   & 0.154   & 18 53 18.50  &   +01 14 59.0 & 1,2 \\
IRAS$\,$18517+0437       & 37.427   & 1.518   & 18 54 13.80  &   +04 41 32.0 & 1   \\
IRAS$\,$18566+0408       & 37.554   & 0.201   & 18 59 09.98  &   +04 12 15.6 & 2   \\
IRAS$\,$19012+0536       & 39.387   &$-$0.140 & 19 03 45.10  &   +05 40 40.0 & 1   \\
G45.07+0.13              & 45.071   & 0.132   & 19 13 22.10  &   +10 50 53.0 & 1   \\
G45.12+0.13              & 45.122   & 0.133   & 19 13 27.80  &   +10 53 37.0 & 1,2 \\
G45.47+0.05              & 45.466   & 0.046   & 19 14 25.60  &   +11 09 26.0 & 1   \\
IRAS$\,$19266+1745$^a$   & 53.032   & 0.117   & 19 28 54.00  &   +17 51 56.0 & 1   \\
IRAS$\,$20051+3435       & 71.892   & 1.312   & 20 07 03.80  &   +34 44 35.0 & 1   \\
G69.54$-$0.98            & 69.540   &$-$0.976 & 20 10 09.10  &   +31 31 34.3 & 2   \\
IRAS$\,$20081+2720$^a$   & 66.153   &$-$3.185 & 20 10 11.50  &   +27 29 06.0 & 1   \\
\hline
\multicolumn{6}{p{12.0cm}}{Two different spectrometer configurations were used in the main survey to observe different molecular transitions (see Table~\ref{tb_config}).} \\
\multicolumn{6}{p{12.0cm}}{$^a$ Observations were conducted toward IRAS coordinates \citep{Sridharan_2002ApJ...566..931S}, which differ by $\sim 30\arcsec$ to $55\arcsec$ with respect to the location of 1.2$\,$mm sources \citep{Beuther_2002ApJ...566..945B}, thus, the observations may be affected by pointing offsets.}
\end{tabular}
\end{threeparttable} 
\end{center}
\end{table*}

\begin{table*}
\begin{center}
\begin{threeparttable} 
\centering
\caption{Spectral Lines Observed in the Main Survey}
\label{tb_config}
\begin{tabular}{lcc ccc} 
\hline
Species  & Rest Frequency& E$_u$/k$_B$ & Chan. Width & Detection$^a$ & Config.$^b$ \\
             & (MHz)	 & (K)    & (\kms)     &          & \\
\hline
CH           & 7275.0040 & 96.7    &  0.13     & N    & 2 \\
CH           & 7325.2030 & 96.7    &  0.13     & N    &  2 \\
CH           & 7348.4190 & 96.7    &  0.06     & E?   &  2 \\
CH           & 7398.6180 & 96.7    &  0.12     & N    &  2 \\
CH$_3$CHO    & 6389.9330 & 8.0     &  0.14     & N    &  1 \\
CH$_3$C$_5$N & 6224.3425 & 8.3     &  0.15     & N    &  1 \\
CH$_3$OH     & 6668.5192 & 49.1    &  0.07     & E, A &  1,2 \\
CH$_3$OH     & 6744.8430 & 803.7   &  0.14     & N    &  2 \\
CH$_3$OH     & 6853.6870 & 353.5   &  0.07     & E?    &  2 \\
CH$_3$OH     & 7283.4490 & 215.9   &  0.06     & E?    &  2 \\
D$_2$CO      & 6096.0682 & 8.4     &  0.15     & N    &  1 \\
H$_2$CS      & 6278.6500 & 23.2    &  0.15     & A    &  1 \\
H99$\alpha$  & 6676.0750 & RRL     &  0.34     & E    &  1 \\
OH           & 6035.0932 & 120.8   &  0.08     & E, A &  1 \\
\hline
\multicolumn{6}{p{11cm}}{Spectroscopic information is from Splatalogue (http://www.cv.nrao.edu/ php/splat/).}\\
\multicolumn{6}{p{11cm}}{$^a$ (E) emission; (A) absorption; (N) non-detection; (?) tentative detection (we report tentative detections only toward IRAS$\,$18566+0408). The tentative detections were not confirmed in our follow-up observations (Table~\ref{tb_followup}), thus it is unclear whether the lines are variable masers or artefacts.}\\
\multicolumn{6}{p{11cm}}{$^b$ The transitions corresponding to Configuration 1 were observed on October 13 and 14, 2008. The Configuration 2 transitions were observed on October 27, 2008.}\\
\end{tabular}
\end{threeparttable} 
\end{center}
\end{table*}

We observed the quasars B1829+290 and B1857+129 to check the pointing performance, telescope gain, system temperature (T$_{sys}$) and half power beam width (HPBW). At a frequency of 6600$\,$MHz, the average system temperature was 30$\,$K, the pointing was better than 15\arcsec (mostly better than 5\arcsec), the HPBW was $\sim 43$\arcsec, and the gain was $\sim 6.0\,$K$\,$Jy$^{-1}$. We checked the OFF-source spectra to look for emission/absorption in the reference position and the dynamic spectra of the ON-source observations to identify radio frequency interference (RFI). We did not find evidence for RFI or other potential bandpass artefacts for the transitions marked as tentative detections in Table~\ref{tb_config}. The T$_{sys} \times $(ON-OFF)/OFF spectra (i.e., the antenna temperature) were calibrated to flux density units (Jy) using the telescope gain. After checking for consistency, the two polarization signals and different scans were averaged, and the final spectra were smoothed to measure the line parameters. The typical RMS was $\sim$5$\,$mJy per spectral window ($\sim 0.1$\kms~typical smoothed channel width; see Table~\ref{tb_config}).

\subsection{Follow-up Observations}

Additional observations with the Arecibo Telescope were conducted on March 17, 2014, to re-observe two of the sources in the sample. We used the WAPP spectrometer to observe 14 different spectral lines observed in two spectrometer setups. The spectrometer setups used in the follow-up observations were different than those used in the main survey (Table~\ref{tb_followup}). The first spectrometer setup was tuned to OH (6.016, 6.030, 6.035, 6.049$\,$GHz), H99$\alpha$, 6.279$\,$GHz H$_2$CS, and 6.668$\,$GHz CH$_3$OH transitions; the second setup was tuned to CH$_3$OH (6.668, 6.744, 6.854, 7.283$\,$GHz) and CH (7.275, 7.325, 7.348, and 7.398$\,$GHz) transitions. Both sets of spectral lines were observed using a 3.125$\,$MHz bandwidth per spectral window, which is equivalent to 125 to 155\kms~for the frequency range of 6.0 to 7.4$\,$GHz of our observations. Each spectral window had 4096 channels, with a channel separation of 0.76$\,$kHz (31 to 38$\,$m$\,$s$^{-1}$). The observations were done in total power ON-source mode. A diode signal was used to calibrate the data to antenna temperature, and the spectra were scaled to flux density units using the telescope gain. All calibration was done using the AOIDL routines provided by the observatory. We observed the calibrator B1857+129 for pointing and system checking. At a frequency of 6650$\,$MHz, the pointing accuracy was better than 5\arcsec, T$_{sys}$ = 32$\,$K, HPBW $\sim \,43$\arcsec, telescope gain $\sim 5.6\,$K$\,$Jy$^{-1}$. After calibration, the data were exported to CLASS\footnote{CLASS is part of the GILDAS software package developed by IRAM.} for baseline subtraction, smoothing, and to measure line parameters. Table~\ref{tb_followup} lists the observed molecules, frequencies, RMS, final channel width after smoothing, and whether a line was detected toward IRAS$\,$18566+0408 and G45.12+0.13. Multiple scans for some configurations were obtained to reduce RMS.

\begin{table*}
\centering
\begin{threeparttable} 
\caption{Follow-up Observations}
\label{tb_followup}
\begin{tabular}{lcc ccc ccc}
\hline
Species & Rest Frequency& E$_u$/k$_B$   & \multicolumn{3}{c}{--------------- IRAS$\,$18566+0408 --------------} & \multicolumn{3}{c}{--------------- G45.12+0.13 ---------------} \\
        & (MHz)          & {(K)}                & {RMS} & {Chan. Width}  & {Detection$^a$} & {RMS} & {Chan. Width} & {Detection$^a$}\\
        &                &                      & {(mJy)} & {(\kms)} &  &{(mJy)} & {(\kms)} & { }\\
\hline
CH	     &  7275.0040 & 96.7  &  3.1    &  0.16  &   N   & 18   &  0.16  &   N   \\
CH	     &  7325.2030 & 96.7  &  RFI    & -      &   -   & RFI  &  -     &   -   \\
CH	     &  7348.4190 & 96.7  &  3.6    &  0.16  &{\bf N}& 18   &  0.16  &   N   \\
CH	     &  7398.6180 & 96.7  & 3.1     &  0.15  & N$^b$ & 18   &  0.15 &   N    \\
CH$_3$OH     &  6668.5192 & 49.1  &  4.9    &  0.03  &   E   &  4.9 &  0.10  &   E, A \\
CH$_3$OH     &  6744.8430 & 803.7 &  3.0    &  0.17  &   N   & 16   &  0.17  &   N   \\
CH$_3$OH     &  6853.6870 & 353.5 &  2.4    &  0.17  &{\bf N}& 15   &  0.17  &   N   \\
CH$_3$OH     &  7283.4490 & 215.9 &  3.2    &  0.16  &{\bf N}& 17   &  0.15  &   N   \\
H99$\alpha$  &  6676.0750 & RRL   &  4.5    &  0.17  &   N   &  3.0 &  0.51  &   E   \\
H$_2$CS      &  6278.6500 & 23.2  &  4.4    &  0.18  &   N   &  4.7 &  0.18  &   N   \\
OH           &  6016.7460 & 120.8 &  5.2    &  0.19  &   N   &  3.3 &  0.38  &   A   \\
OH           &  6030.7485 & 120.8 &  4.5    &  0.19  &   N   &  7.1 &  0.08 &   E, A \\
OH           &  6035.0932 & 120.8 &  4.6    &  0.19  &   E   & 6.7 &  0.08 &   E, A \\
OH           &  6049.0840 & 120.8 &  3.5    &  0.38  &  N$^b$&  5.4 &  0.19  &   N   \\
\hline
\multicolumn{9}{p{16cm}}{Radio continuum was not measured because the observations were conducted in total power ON-source mode. Spectroscopic information is from Splatalogue (http://www.cv.nrao.edu/php/splat/).}\\
\multicolumn{9}{p{16cm}}{$^a$ (-) spectral window corrupted by RFI; (E) emission; (A) absorption; (N) non-detection. We highlight with boldface ({\bf N}) transitions that were tentatively detected in the main survey (Table~\ref{tb_config}) but not detected in the follow-up observations.}\\
\multicolumn{9}{p{16cm}}{$^b$ Two narrow variable lines at 96.2 and 98.3\kms~were detected in most IRAS$\,$18566+0408 7.398$\,$GHz CH scans, most likely caused by low-level RFI. A weak ($\sim 10\,$mJy) 6.049$\,$GHz OH feature at 60\kms~was also detected toward IRAS$\,$18566+0408, but it is most likely a bandpass artefact.}\\
\end{tabular}
\end{threeparttable} 
\end{table*}

\begin{table*}
\begin{center}
\begin{threeparttable} 
\centering
\small
\setlength{\tabcolsep}{2.0pt}
\caption{Detection Summary}
\label{tb_sources_summary}
\begin{tabular}{l ccc c cc c c}
\hline
    Source           & \multicolumn{3}{c}{--------------- OH ---------------} & {-- H$_2$CS --}    & \multicolumn{2}{c}{------ CH$_3$OH ------} & {--- CH ---} & {-- H99$\alpha$ --}\\
	         & 6.016$\,$GHz & 6.031$\,$GHz & 6.035$\,$GHz & 6.279$\,$GHz& 6.668$\,$GHz & 7.283$\,$GHz & 7.348$\,$GHz & 6.676$\,$GHz \\
\hline
IRAS$\,$18472$-$0022 & ... 	        & ...          & N 	          & N 	      & N 	   & ... 	        & ... 	     & N\\
G34.26+0.15          & ... 	        & ...          & E,A 	  & A 	      & E,A 	   & N 	        & N 	     & E\\
IRAS$\,$18517+0437   & ... 	        & ...          & N 	          & N 	      & E 	   & ... 	        & ... 	     & N\\
IRAS$\,$18566+0408   & N 	        & N            & E 	          & N 	      & E 	   & E? 	        & E? 	     & N\\
IRAS$\,$19012+0536   & ... 	        & ...          & N 	          & N 	      & E 	   & ... 	        & ... 	     & N\\    
G45.07+0.13          & ... 	        & ...          & E,A 	  & N 	      & E 	   & ... 	        & ... 	     & E\\ 
G45.12+0.13          & A 	        & E,A          & E,A 	  & N 	      & E?,A 	   & N 	        & N 	     & E\\ 
G45.47+0.05          & ... 	        & ...          & E 	          & N 	      & E,A 	   & ... 	        & ... 	     & E\\ 
IRAS$\,$19266+1745   & ... 	        & ...          & N 	          & N 	      & E 	   & ... 	        & ... 	     & N\\   
IRAS$\,$20051+3435   & ... 	        & ...          & N 	          & N 	      & N 	   & ... 	        & ... 	     & N\\
G69.54$-$0.98        & ... 	        & ...          & ... 	  & ... 	      & E,A 	   & N 	        & N 	     & ...\\
IRAS$\,$20081+2720   & ... 	        & ...          & N 	          & N 	      & N 	   & ... 	        & ... 	     & N\\    
\hline
\multicolumn{9}{p{14.5cm}}{Summary of transitions with at least one detection. (E) emission; (A) absorption; (N) non-detection; (...) not observed; (?) tentative detection. }\\
\end{tabular}
\end{threeparttable} 
\end{center}
\end{table*}

\section{Results} \label{Results}

We report detection of 6.7$\,$GHz CH$_3$OH emission toward nine regions (although the line toward G45.12+0.13 could be due to a bright maser in the sidelobe of the telescope, see below), and 6.7$\,$GHz CH$_3$OH absorption toward four sources (3 new detections: G45.12+0.13, G45.47+0.05, and G69.54$-$0.98). We also report 6.035$\,$GHz OH emission toward five sources (G45.07+0.13 being a new detection to our knowledge); H99$\alpha$ emission toward four sources; and the first detection of 6.016$\,$GHz and 6.030$\,$GHz OH lines toward G45.12+0.13. A summary of detections and non-detections of the different spectral lines observed in this project is listed in Tables~\ref{tb_config}, \ref{tb_followup} and \ref{tb_sources_summary}. Table~\ref{tb_CH3OH-6668} lists the line parameters of the nine sources with detection of CH$_3$OH lines at 6.7$\,$GHz. IRAS$\,$18517+0437 shows the highest 6.7$\,$GHz CH$_3$OH maser flux density (212$\,$Jy); the weakest 6.7$\,$GHz CH$_3$OH line was detected toward IRAS$\,$19012+0536 (20$\,$mJy); and the strongest absorption was found toward G34.26+0.15 ($-$318$\,$mJy). Table~\ref{tb_OH6035} lists detections of the 6.035$\,$GHz OH line toward five high-mass star forming regions. The highest flux density was 7.51$\,$Jy toward G45.47+0.05 and the strongest absorption was $-$129$\,$mJy toward G45.12+0.13. The H99$\alpha$ line parameters are listed in Table~\ref{tb_H99a}. Table~\ref{tb_tentdetec} shows other detections and tentative detections. Three tentative lines were found toward IRAS$\,$18566+0408: 6.854$\,$GHz CH$_3$OH, 7.283$\,$GHz CH$_3$OH and 7.348$\,$GHz CH; however, these lines were not confirmed in our follow-up observations (Table~\ref{tb_followup}). It is unclear whether the tentative detections are variable lines or artefacts (see discussion in section \ref{Tentative}). 

In Tables~\ref{tb_CH3OH-6668}, \ref{tb_OH6035}, and \ref{tb_tentdetec} we list the peak channel flux density, peak channel velocity (channel width listed as uncertainty), line-width above 3$\sigma$\footnote{In case of overlapping lines, the linewidth above 3$\sigma$ is measured up to the local minimum of the overlapping lines. For example, if two lines overlap, two linewidths are reported, one from the first channel above 3$\sigma$ to the local minimum of the overlapping lines, and the other linewidth from the local minimum to the last channel above 3$\sigma$ of the second line.} (2 times the channel width listed as uncertainty), and integrated flux density. In a few cases where the lines were broad and approximately symmetric, the line parameters were instead obtained from Gaussian fits as indicated in table notes. The spectra of the detections and tentative detections are presented in Figures~\ref{fCH3OH} to \ref{ftenative_detections}. To facilitate the use of our data, we also provide as supplemental electronic material the ASCII files used to generate the figures.

A potential problem in a targeted high-sensitivity survey such as ours is the possibility of sources at the edge of the primary beam and/or contamination by unrelated sources outside of the primary beam, i.e., spurious detections of nearby bright sources in the sidelobes. As pointed out by \cite{Breen_2015MNRAS.450.4109B}, such contamination is possible in the case of weak lines detected with Arecibo. A source in our sample that could suffer from such contamination is the 6.7$\,$GHz CH$_3$OH maser in G45.12+0.13 (Figure~\ref{f_CH3OH_G45}), as we detected a strong maser toward the source G45.07+0.13 (Table~\ref{tb_CH3OH-6668}; Figure~\ref{fCH3OH}) at the same peak velocity (57.83$\pm$0.07\kms~for G45.07+0.13 vs 57.89$\pm$0.07 \kms~for G45.12+0.13). The angular separation between G45.07+0.13 and G45.12+0.13 is 3\arcmin, which is significantly greater than the HPBW of the telescope at 6.6$\,$GHz ($\sim 43$\arcsec) as well as greater than the location of the first sidelobe of the telescope ($\sim 1.2$\arcmin; see beam pattern in \citealt{2007ApJ...656..255P}). However, the flux density ratio between the strong maser in G45.07+0.13 and the weak maser in G45.12+0.13 is quite large ($\sim 700$), and thus, we cannot rule out contamination from a higher-order sidelobe. A cross scan with the Arecibo Telescope or interferometric observations are needed to clarify the nature of the weak emission line toward G45.12+0.13.

Contamination by bright 6.7$\,$GHz CH$_3$OH masers outside the primary beam for the other sources in our sample seems unlikely, as we found very similar spectral profiles for the other sources in common with the Methanol MultiBeam (MMB) survey \citep{Breen_2015MNRAS.450.4109B}. Also, previous interferometric observations of some of the sources show very similar spectral profiles to those reported here (e.g.,  \citealt{Xu_2009A&A...507.1117X}, \citealt{araya10}, \citealt{2011ApJ...730...55P}). However, we emphasize that cross-scans and/or interferometric observations are needed to completely rule out contamination in the case of weak lines.

\begin{table*}
\begin{threeparttable} 
\centering
\caption{6.7$\,$GHz CH$_3$OH Line Parameters}
\label{tb_CH3OH-6668}
\begin{tabular}{lcccccc} 
\hline
Source       & S$_{\nu, cont}$ & RMS  & S$_{\nu}$ & V$_{LSR}$    & Width  & $\int S_\nu dv$ \\
                   & (Jy)   &(mJy) & (Jy)    & (\kms)      & (\kms) & (Jy\kms) \\
\hline
G34.26+0.15        & 4.0   & 6.8  & 0.105 & 55.29(0.07) & 1.0(0.1) & 0.063\\
	           & 4.0   & 6.8  & 1.42  & 56.80(0.07) & 1.5(0.1) & 0.835\\
                   & 4.0   & 6.8  & 22.9  & 57.69(0.07) & 0.7(0.1) & 8.85\\
                   & 4.0   & 6.8  & 22.0  & 57.90(0.07) & 1.0(0.1) & 7.71\\
                   & 4.0   & 6.8  & 0.148 & 59.61(0.07) & 0.7(0.1) & 0.067 \\
                   & 4.0   & 6.8  & 2.04  & 60.64(0.07) & 0.7(0.1) & 0.738 \\
                   & 4.0   & 6.8  & 1.51  & 61.19(0.07) & 0.4(0.1) & 0.458 \\
                   & 4.0   & 6.8  & 0.664 & 61.60(0.07) & 0.6(0.1) & 0.197 \\
		   & 4.0   & 6.8  & $-$0.318 & 62.06(0.07) & [3.1, 5.8]$^a$ & ... \\
IRAS$\,$18517+0437 & $<$0.1 & 3.2  & 212  & 41.22(0.07) & 2.6(0.1)	& 126\\
                   & $<$0.1 & 3.2  & 0.239    & 42.80(0.07) & 0.8(0.1)	& 0.118\\
                   & $<$0.1 & 3.2  & 2.03    & 45.20(0.07) & 2.1(0.1)	& 1.89\\
                   & $<$0.1 & 3.2  & 2.93    & 46.02(0.07) & 1.6(0.1)	& 1.44\\
                   & $<$0.1 & 3.2  & 0.164    & 47.64(0.07) & 0.6(0.1)	& 0.084\\
                   & $<$0.1 & 3.2  & 0.380    & 48.39(0.07) & 1.0(0.1)	& 0.209\\
                   & $<$0.1 & 3.2  & 0.507    & 49.32(0.07) & 0.8(0.1)	& 0.204\\
                   & $<$0.1 & 3.2  & 0.140    & 49.91(0.07) & 1.1(0.1)	& 0.124\\
                   & $<$0.1 & 3.2  & 1.34    & 51.44(0.07) & 1.2(0.1) 	& 0.627\\
IRAS$\,$18566+0408 & $<$0.1 & 4.5  & 0.367    & 78.48(0.07) & 0.7(0.1) & 0.139\\
	           & $<$0.1 & 4.5 & 2.19    & 79.72(0.07) & 1.6(0.1) & 1.43\\
	           & $<$0.1 & 4.5 & 4.89    & 83.77(0.07) & 1.0(0.1) & 1.87\\
	           & $<$0.1 & 4.5 & 2.38    & 84.59(0.07) & 0.6(0.1) & 1.29\\
	           & $<$0.1 & 4.5 & 3.24    & 84.94(0.07) & 0.8(0.1) & 1.41\\
	           & $<$0.1 & 4.5 & 0.444    & 85.61(0.07) & 0.4(0.1) & 0.181\\
	           & $<$0.1 & 4.5 & 2.03    & 86.38(0.07) & 1.5(0.1) & 1.30\\
	           & $<$0.1 & 4.5 & 0.351    & 87.88(0.07) & 0.9(0.1) & 0.187\\
                   & $^b$   & 4.9  & 0.175   & 78.48(0.03) & 0.7(0.1) & 0.070\\
                   & $^b$   & 4.9  & 1.55   & 79.71(0.03) & 1.5(0.1) & 1.08\\
                   & $^b$   & 4.9  & 4.25   & 83.76(0.03) & 0.9(0.1) & 1.53\\
                   & $^b$   & 4.9  & 2.04   & 84.58(0.03) & 0.7(0.1) & 0.981\\
                   & $^b$   & 4.9  & 3.57   & 84.96(0.03) & 0.6(0.1) & 1.25\\
                   & $^b$   & 4.9  & 0.567  & 85.61(0.03) & 0.6(0.1) & 0.228\\
                   & $^b$   & 4.9  & 1.78   & 86.40(0.03) & 1.1(0.1) & 1.02\\
                   & $^b$   & 4.9  & 0.556   & 87.84(0.03) & 0.8(0.1) & 0.216\\
IRAS$\,$19012+0536 & $<$0.1 & 3.4  & 0.410    & 58.52(0.07) & 0.7(0.1)	& 0.185\\
                   & $<$0.1 & 3.4  & 0.542    & 58.92(0.07) & 0.6(0.1)	& 0.236\\
                   & $<$0.1 & 3.4  & 0.253    & 59.61(0.07) & 0.5(0.1)	& 0.129\\
                   & $<$0.1 & 3.4  & 1.16    & 60.43(0.07) & 1.3(0.1)	& 0.662\\
                   & $<$0.1 & 3.4  & 0.026    & 61.19(0.07) & 0.3(0.1)	& 0.008\\
                   & $<$0.1 & 3.4  & 0.286    & 61.61(0.07) & 0.9(0.1)	& 0.113\\
                   & $<$0.1 & 3.4  & 0.049    & 62.36(0.07) & 0.3(0.1)	& 0.020\\
                   & $<$0.1 & 3.4  & 0.108    & 62.98(0.07) & 0.8(0.1)	& 0.059\\
                   & $<$0.1 & 3.4  & 0.038    & 63.46(0.07) & 0.5(0.1)	& 0.016\\
                   & $<$0.1 & 3.4  & 0.078    & 64.76(0.07) & 0.8(0.1)	& 0.042\\
                   & $<$0.1 & 3.4  & 0.052    & 65.45(0.07) & 0.6(0.1)	& 0.024\\
                   & $<$0.1 & 3.4  & 0.032    & 66.00(0.07) & 0.4(0.1)	& 0.013\\
                   & $<$0.1 & 3.4  & 0.032    & 66.41(0.07) & 0.9(0.1)	& 0.016\\ 
                   & $<$0.1 & 3.4  & 0.448    & 68.60(0.07) & 1.1(0.1)	& 0.311\\
                   & $<$0.1 & 3.4  & 0.321    & 69.43(0.07) & 0.7(0.1)	& 0.193\\
                   & $<$0.1 & 3.4  & 0.216    & 69.84(0.07) & 0.4(0.1)	& 0.071\\
                   & $<$0.1 & 3.4  & 0.057    & 70.26(0.07) & 0.3(0.1)	& 0.018\\
                   & $<$0.1 & 3.4  & 0.058    & 70.63(0.07) & 0.6(0.1)	& 0.027\\
                   & $<$0.1 & 3.4  & 0.054    & 71.83(0.07) & 0.9(0.1)	& 0.027\\
                   & $<$0.1 & 3.4  & 0.020    & 72.79(0.07) & 0.4(0.1)	& 0.007\\
                   & $<$0.1 & 3.4  & 0.053    & 75.47(0.07) & 0.5(0.1)	& 0.002\\ 
\hline
\end{tabular}
\end{threeparttable} 
\end{table*}

\begin{table*}
\begin{threeparttable} 
\centering
\addtocounter{table}{-1}
\caption{Continued}
\label{tb_CH3OH-6668}
\begin{tabular}{lccccccccccc} 
\hline
Source          & S$_{\nu, cont}$ &  RMS    & S$_{\nu}$ & V$_{LSR}$    &  Width     & $\int S_\nu dv$ \\
                & (Jy)         & (mJy)  & (Jy)     & (\kms)      & (\kms)     & (Jy\kms)       \\
\hline
G45.07+0.13     & 0.4	       & 3.5	& 0.029	   & 55.97(0.07) & 0.5(0.1) & 0.012\\
                & 0.4	       & 3.5	& 41.5    & 57.83(0.07) & 2.2(0.1) & 19.3\\
                & 0.4	       & 3.5	& 0.397	   & 59.75(0.07) & 0.9(0.1) & 0.151\\
G45.12+0.13     & 4.1	       & 6.3	& 0.063     & 57.89(0.07) & 0.8(0.1) & 0.032 \\
$^{c}$          & 4.1	       & 6.3	& $-$0.026  & 59.3(0.2)   & 3.3(0.3) & $-$0.091 \\ 
$^{c, d}$        & 4.9	       & 8.0	& $-$0.024  & 59.5(0.2)   & 3.0(0.4) & $-$0.078 \\  
                & $^{b}$       & 5.6    & 0.046     & 57.8(0.1)   & 0.7(0.2) & 0.021 \\
$^{c}$          & $^{b}$       & 5.6    & $-$0.024  & 59.6(0.1)   & 3.8(0.3) & $-$0.096 \\ 
G45.47+0.05     & 0.4	       & 3.4	& 1.04	   & 56.04(0.07) & 0.8(0.1) & 0.498\\ 
                & 0.4	       & 3.4	& 0.880	   & 56.52(0.07) & 0.8(0.1) & 0.477\\
                & 0.4	       & 3.4	& 0.960	   & 57.48(0.07) & 0.8(0.1) & 0.470\\
                & 0.4	       & 3.4	& 0.860	   & 58.17(0.07) & 0.9(0.1) & 0.422\\
                & 0.4	       & 3.4	& 0.150	   & 59.13(0.07) & 0.7(0.1) & 0.063\\
                & 0.4	       & 3.4	& 0.060	   & 59.61(0.07) & 0.4(0.1) & 0.019\\
                & 0.4	       & 3.4	& 0.040	   & 66.27(0.07) & 2.0(0.1) & 0.075\\
                & 0.4	       & 3.4	& $-$0.030  & 63.25(0.07) & 3.4(0.1) & $-$0.056\\
IRAS$\,$19266+1745 & $<$0.1    & 3.6	& 0.930	   & 10.14(0.07) & 0.8(0.1) & 0.421\\ 
G69.54$-$0.98	& 0.1	       & 3.8	& 13.0      & 0.00(0.07)  & 1.2(0.1) &  6.37   \\
	        & 0.1	       & 3.8	&  2.28     & 1.16(0.07)  & 1.2(0.1) &  1.19   \\
	        & 0.1	       & 3.8	&  0.115    & 2.47(0.07)  & 0.9(0.1) &  0.063  \\
	        & 0.1	       & 3.8	& 57.1      & 14.68(0.07) & 1.8(0.1) & 20.8    \\
	        & 0.1	       & 3.8	&  0.469    & 15.64(0.07) & 0.5(0.1) &  0.131  \\
	        & 0.1	       & 3.8	& $-$0.018  & 11.87(0.07) & [1.6, 5.2]$^a$ & ... \\
\hline
\multicolumn{7}{p{12cm}}{${^a}$~A linewidth range ([minimum, maximum]) is given as overlapping maser lines preclude exact measurement. Integrated flux density could not be reliably measured because the absorption line is blended with a strong maser. The absorption was fit and removed from the spectrum to measure the line parameters of the masers.}\\
\multicolumn{7}{p{12cm}}{$^{b}$ Data from March 2014 observations (see Table~\ref{tb_followup}; radio continuum was not measured because the observations were conducted in total power ON-source mode).} \\
\multicolumn{7}{p{12cm}}{$^c$ Line parameters are from a Gaussian fit, $1\sigma$ statistical errors from the fit are listed. The FWHM is listed as width.}\\ 
\multicolumn{7}{p{12cm}}{$^{d}$~G45.12+0.13 was also observed on 2008 October 27.}\\
\end{tabular}
\end{threeparttable} 
\end{table*}

\begin{table*}
\begin{threeparttable} 
\centering
\caption{Line Parameters of 6.035$\,$GHz OH}
\label{tb_OH6035}
\begin{tabular}{lcccccc} 
\hline
Source        & S$_{\nu, cont}$ & RMS   & S$_{\nu}$& V$_{LSR}$       & Width         & $\int S_\nu dv$\\
              & (Jy)         & (mJy) & (Jy)    & (\kms)         & (\kms)        & (Jy\kms) \\
\hline
G34.26+0.15   & 4.3         & 8.1   &  0.282 & 55.84(0.08)    & 1.0(0.2)    &  0.122  \\
	      & 4.3         & 8.1   &  0.027 & 56.90(0.08)    & 0.4(0.2)    &  0.007  \\
	      & 4.3         & 8.1   &  0.351 & 58.19(0.08)    & 0.8(0.2)    &  0.116  \\
              & 4.3         & 8.1   &  0.465 & 58.49(0.08)    & 0.7(0.2)    &  0.163  \\
	      & 4.3         & 8.1   &  0.171 & 59.63(0.08)    & 0.5(0.2)    &  0.050  \\
	      & 4.3         & 8.1   &  0.062 & 60.24(0.08)    & 0.4(0.2)    &  0.019  \\
	      & 4.3         & 8.1   &  0.352 & 60.92(0.08)    & 0.9(0.2)    &  0.155  \\
	      & 4.3         & 8.1   &  0.119 & 62.06(0.08)    & 0.4(0.2)    &  0.030  \\
	      & 4.3         & 8.1   &  0.163 & 62.51(0.08)    & 0.5(0.2)    &  0.048  \\
	      & 4.3         & 2.9$^a$& $-$0.062& 50.8(1.1)    & [57, 80]    & ...    \\ 
IRAS$\,$18566+0408 & $^b$   & 4.6   & 0.014  & 85.8(0.2)      & 0.6(0.4)     &  0.007 \\
G45.07+0.13   & 0.4         & 3.1   & 0.232  & 55.85(0.08)    & 0.8(0.2)  & 0.098\\
	      & 0.4         & 3.1   & 0.128  & 57.59(0.08)    & 1.0(0.2)  & 0.065\\ 
	      & 0.4         & 3.1   & 0.018  & 58.19(0.08)    & 0.5(0.2)  & 0.008\\
	      & 0.4         & 3.1   & 0.015  & 60.69(0.08)    & 0.4(0.2)  & 0.007\\ 
	      & 0.4         & 3.1   & 0.086  & 62.75(0.08)    & 0.6(0.2)  & 0.026\\ 
	      & 0.4         & 3.1   & $-$0.021 & 50.84(0.08)  & 10.0(0.2) & $-$0.089\\ 
G45.12+0.13   & 4.1$^c$     & 5.5   & $-$0.024 & 51.8(0.3)    & 1.6(0.6)  & $-$0.041\\ 
              & 4.1         & 5.5   & 2.20   & 53.87(0.08)    & 0.9(0.2)  & 1.13 \\
              & 4.1         & 5.5   & 2.54   & 54.32(0.08)    & 1.0(0.2)  & 1.42 \\
              & 4.1         & 5.5   & 1.16   & 55.54(0.08)    & 0.8(0.2)  & 0.475 \\
              & 4.1$^c$     & 5.5   & $-$0.111 & 58.24(0.09) & 3.3(0.2)  & $-$0.390\\
	      &$^{b,c}$     & 7.1   & $-$0.035 & 51.9(0.3)    & 1.4(0.7)  & $-$0.051 \\
	      &$^b$         & 7.1   & 2.54   & 53.87(0.08)    & 1.9(0.2)  & 2.495 \\
	      &$^b$         & 7.1   & 1.75   & 55.49(0.08)    & 0.8(0.2)  & 0.680 \\
	      &$^{b,c}$     & 7.1   & $-$0.129 & 58.2(0.1)    & 3.5(0.4)  & $-$0.484 \\ 
G45.47+0.05   & 0.5          & 3.7   & 0.018   & 60.09(0.08)	& 0.4(0.2)	& 0.004\\
	      & 0.5          & 3.7   & 0.040    & 60.62(0.08)	& 0.3(0.2)	& 0.012\\
	      & 0.5          & 3.7   & 0.013    & 61.15(0.08)	& 0.4(0.2)	& 0.004\\
	      & 0.5          & 3.7   & 0.049    & 61.60(0.08)	& 0.5(0.2)	& 0.017\\
	      & 0.5          & 3.7   & 0.033    & 61.91(0.08)	& 0.5(0.2)	& 0.014\\
	      & 0.5          & 3.7   & 0.174    & 62.82(0.08)	& 0.7(0.2)	& 0.063\\
	      & 0.5          & 3.7   & 0.752    & 63.28(0.08)	& 0.6(0.2)	& 0.323\\
	      & 0.5          & 3.7   & 0.510    & 63.67(0.08)	& 0.5(0.2)	& 0.228\\
	      & 0.5          & 3.7   & 0.902    & 64.41(0.08)	& 0.6(0.2)	& 0.445\\
	      & 0.5          & 3.7   & 3.84    & 65.02(0.08)	& 0.8(0.2)	& 1.96\\
	      & 0.5          & 3.7   & 3.10    & 65.86(0.08)	& 0.7(0.2)	& 1.66\\
	      & 0.5          & 3.7   & 7.51    & 66.31(0.08)	& 1.3(0.2)	& 3.76\\
	      & 0.5          & 3.7   & 3.21    & 68.14(0.08)	& 1.9(0.2)	& 2.66\\ 
\hline
\multicolumn{7}{p{12cm}}{$^a$ Spectra smoothed to a channel width of 1.1\kms~to measure the peak flux density and velocity of a weak and broad absorption line. Because of the broad asymmetric absorption profile and overlapping bright masers, a linewidth range is given ([minimum, maximum]), and the integrated flux density was not reliably measured. The absorption was fit and removed from the spectrum to measure the line parameters of the masers.}\\
\multicolumn{7}{p{12cm}}{$^b$ Data from March 2014 observations (see Table \ref{tb_followup}; radio continuum was not measured because the observations were conducted in total power ON-source mode).}\\
\multicolumn{7}{p{12cm}}{$^{c}$ Line parameters from Gaussian fit. FWHM reported as line width. Two absorption lines fitted with independent Gaussians in this spectrum, however, it is likely that the absorption is a single broad line blended with masers.}
\end{tabular}
\end{threeparttable} 
\end{table*}

\begin{table*}
\begin{threeparttable} 
\centering
\caption{H99$\alpha$ Line Parameters}
\label{tb_H99a}
\begin{tabular}{lccccccccccc} 
\hline
Source      & S$_{\nu, cont}$ & RMS   & S$_{\nu}$ & V$_{LSR}$    & FWHM        & $\int S_\nu dv$\\
            & (Jy)         & (mJy) & (Jy)    & (\kms)      & (\kms)      & (Jy\kms)\\
\hline
G45.07+0.13 & 0.4         & 2.2  & 0.012    & 55.5(0.5) & 24.9(1.1) & 0.313\\
G45.47+0.05 & 0.5         & 2.6  & 0.022    & 59.8(0.3) & 31.7(0.7) & 0.733\\
\hline
\multicolumn{7}{p{10cm}}{Line parameters obtained from Gaussian fits; 1$\sigma$ statistical errors from the fit listed as uncertainties. As shown in Figure~\ref{RRL}, H99$\alpha$ lines were also detected toward G34.26+0.15 and G45.12+0.13, however, it was not possible to subtract the baseline as the lines were very board with respect to the bandpass, thus, line parameters were not reliably measured. In the case of G34.26+0.15, \citet{Sewilo_2004ApJ...605..285S} showed that the broad RRL profile is the convolution of multiple velocity components. Line parameters of G45.12+0.13 RRLs are reported in \citet{Araya_2002ApJS..138...63A} and \citet{Wood_1989ApJS...69..831W}.}
\end{tabular}
\end{threeparttable} 
\end{table*}

\begin{table*}
\begin{threeparttable} 
\centering
\caption{Other Detections and Tentative Detections}
\label{tb_tentdetec}
\begin{tabular}{lccccccccccc} 
\hline
Source                & Molecule$^a$ & Frequency & RMS   & S$_{\nu}$     &  V$_{LSR}$    & Width      & $\int S_\nu dv$ \\
                      &          & (MHz)     & (mJy) & (Jy)         &  (\kms)     & (\kms)     & (Jy\kms)        \\
\hline
IRAS$\,$18566+0408 (?)& CH$_3$OH & 6853.6870 & 5.0   & 0.017        & 83.27(0.07) & 0.6(0.2) & 0.004 \\
IRAS$\,$18566+0408 (?)& CH$_3$OH & 7283.4490 & 4.6   & 0.021        & 90.41(0.06) & 0.4(0.2) & 0.003 \\
                      &          &           & 4.6   & 0.019        & 91.98(0.06) & 0.1(0.2) & 0.005 \\
IRAS$\,$18566+0408 (?)& CH       & 7348.4190 & 5.5   & 0.020        & 88.74(0.06) & 0.4(0.2) & 0.004 \\
G45.12+0.13$^{b,c}$   & OH       & 6016.7460 & 2.5   & $-$0.026     & 55.8(0.2)   & 7.8(0.4)   & $-$0.214\\
G45.12+0.13$^{b,c,d}$ & OH       & 6030.7485 & 6.3   & $-$0.029     & 52.4(0.1)   & 1.9(0.3)   & $-$0.059 \\
$^{b}$               &          &           & 6.3   & 0.214        & 53.90(0.08) & 2.0(0.2)   &  0.245 \\
$^{b,c,d}$            &          &           & 6.3   & $-$0.100     & 57.89(0.03) & 4.32(0.08) & $-$0.460\\
\hline
\multicolumn{8}{p{15cm}}{Tentative detections are marked with question marks (?).} \\
\multicolumn{8}{p{15cm}}{$^a$ We also detected a line of thioformaldehyde (H$_2$CS) at 6.279$\,$GHz toward G34.26+0.15 [$S_\nu = 16(1)\,$mJy, $V_{LSR} = 62.6(0.2)$\kms, $FWHM = 3.9(0.4)$\kms; parameters from a Gaussian fit]; a discussion of this detection including follow-up observations will be the topic of a future article.} \\
\multicolumn{8}{p{15cm}}{$^{b}$ March 2014 observations, see Table \ref{tb_followup}.} \\
\multicolumn{8}{p{15cm}}{$^{c}$ Line parameters from Gaussian fit. FWHM reported as line width.}\\
\multicolumn{8}{p{15cm}}{$^{d}$ Two absorption lines fitted with independent Gaussians, however, it is likely that the absorption is a single broad line blended with masers.} \end{tabular}
\end{threeparttable} 
\end{table*}


\begin{figure*}
\includegraphics[width=15cm]{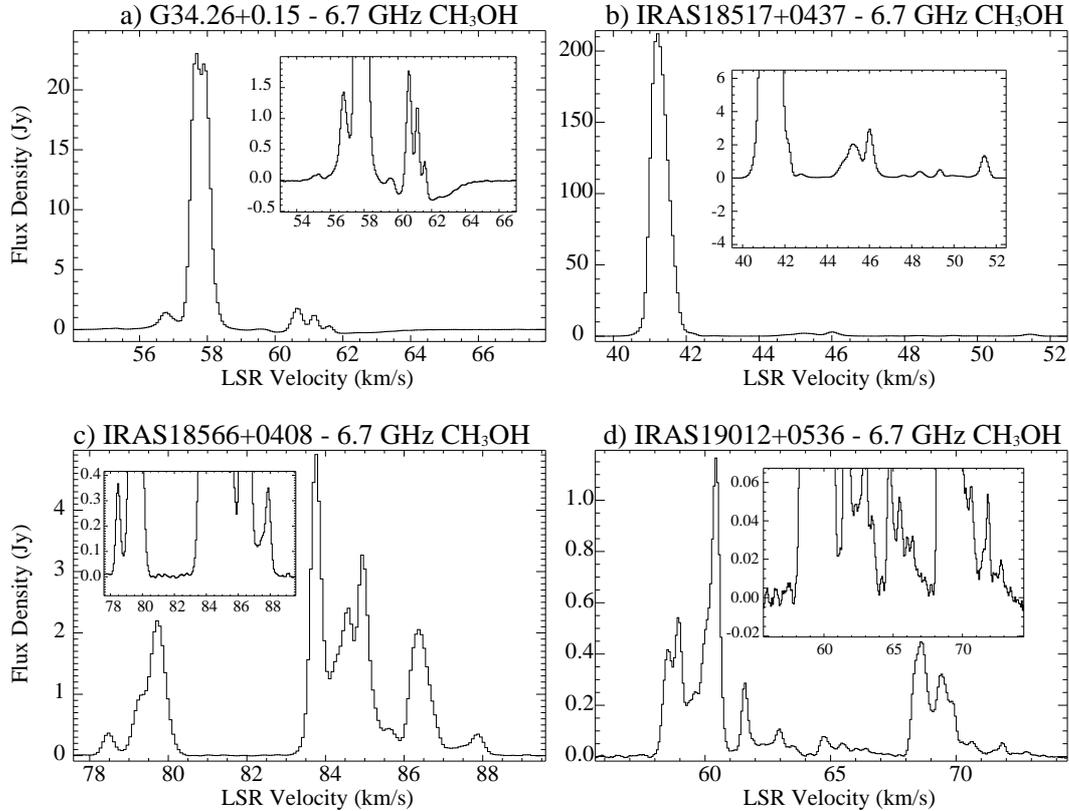}
\vspace*{-7.5cm}\caption{Detection of CH$_3$OH lines toward eight sources (see Table~\ref{tb_CH3OH-6668}; the G45.12$+$0.13 spectra is shown in Figure~\ref{f_CH3OH_G45}). Insets highlight weak masers and/or absorption. The data shown in the figure are available as supplemental online material.} 
\label{fCH3OH}
\end{figure*}

\begin{figure*}
\addtocounter{figure}{-1}
\includegraphics[width=15cm]{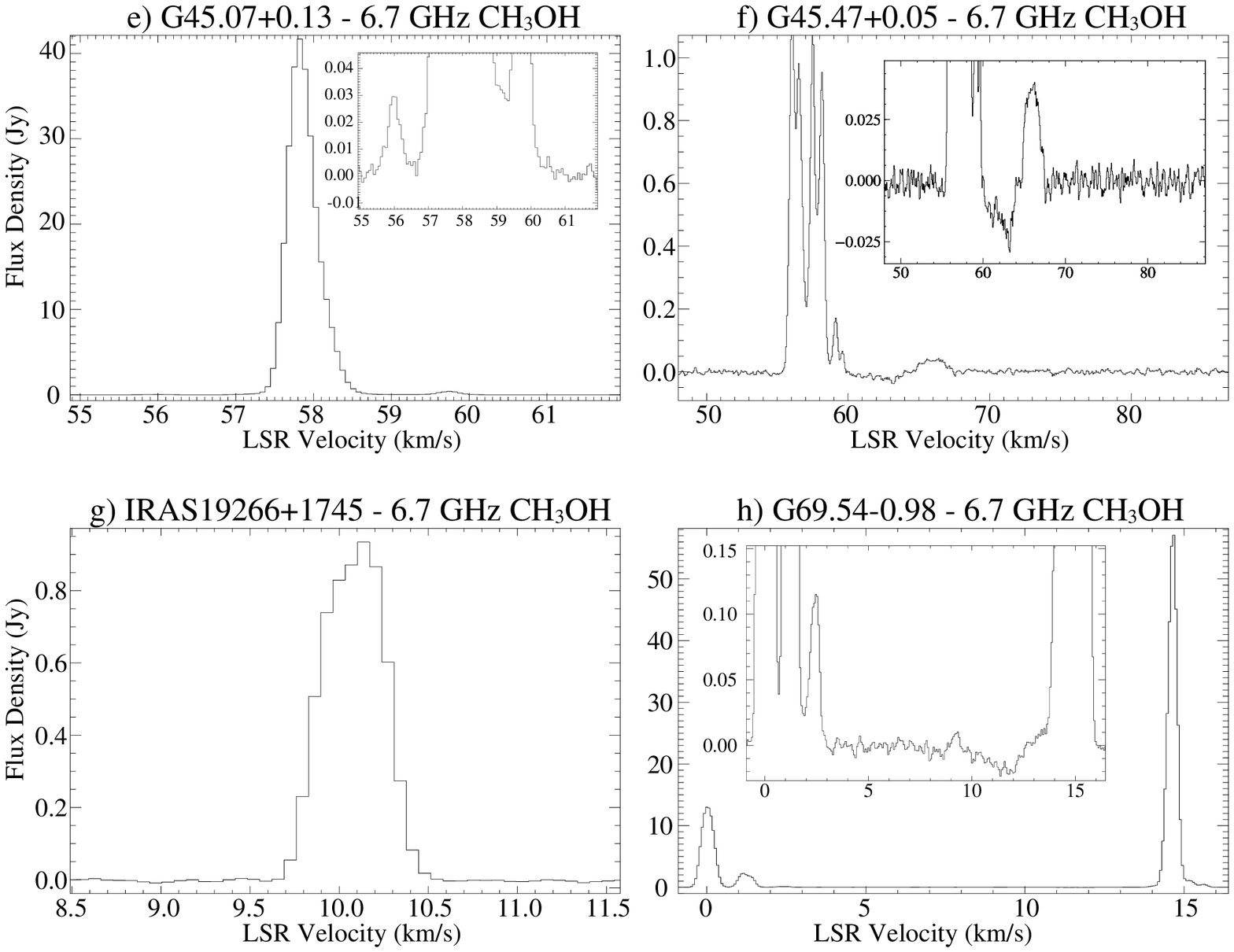}
\vspace*{-7.5cm}\caption{ (continued). }
\label{fCH3OHCont}
\end{figure*}

\begin{figure*}
\includegraphics[width=12cm]{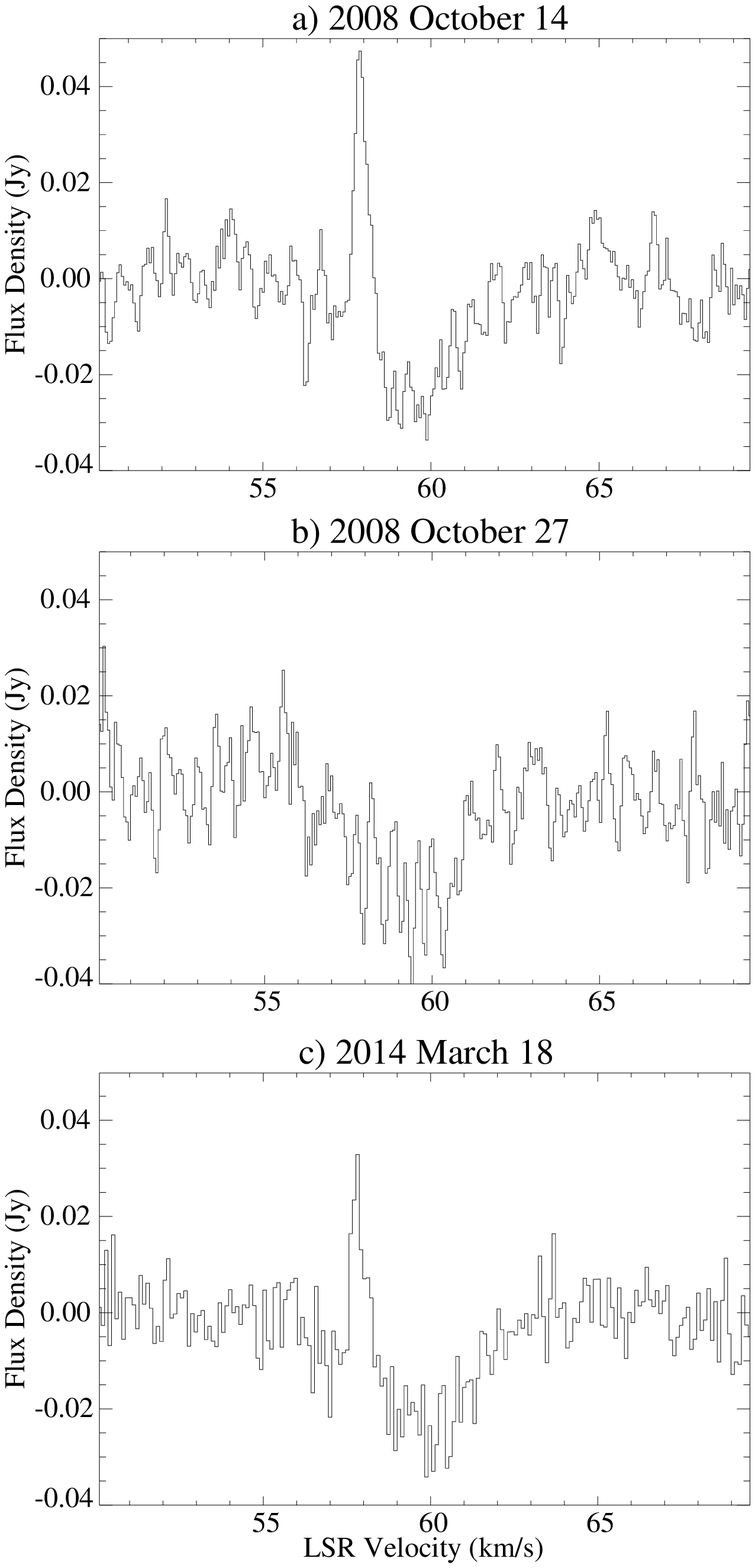}
\caption{CH$_3$OH spectra of the 6.7$\,$GHz transition toward G45.12+0.13 obtained in three epochs: a) 2008 October 14; b) 2008 October 27; c) 2014 March 18. Note the detection of a variable recurrent maser; as explained in section \ref{Results}, the line could be caused by detection of the strong maser in G45.07+0.13 in a sidelobe of the telescope. The data shown in the figure are available as supplemental online material.}
\label{f_CH3OH_G45}
\end{figure*}

\begin{figure*}
\includegraphics[angle = 90,width=15cm]{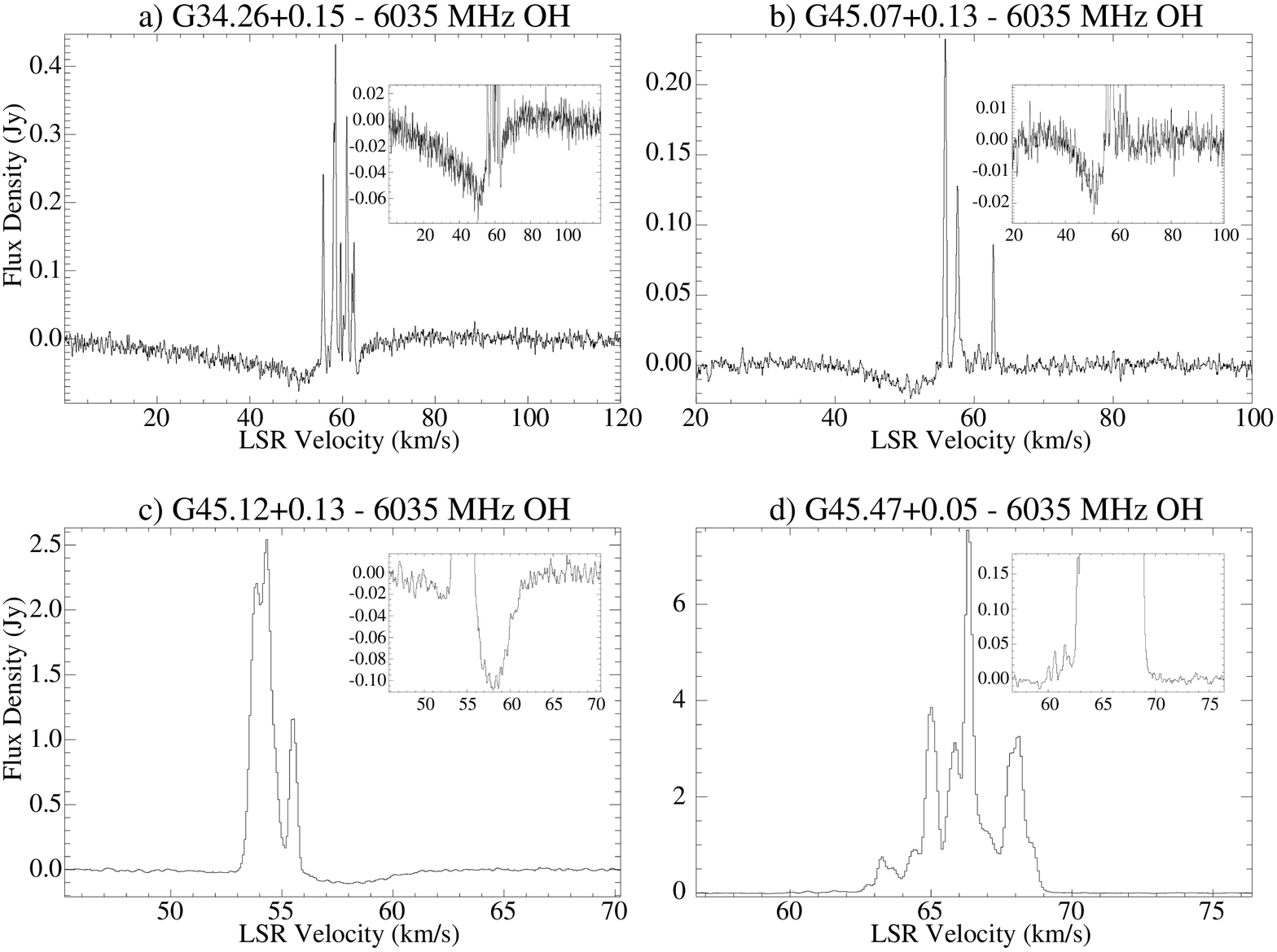}
\caption{Detection of excited OH lines at 6.035$\,$GHz toward four sources (see Table~\ref{tb_OH6035}). Emission lines were detected in all cases; absorption was also found toward three regions. The inset panels show zoom-in views of the spectral lines. The data shown in the figure are available as supplemental online material.}
\label{fig_6035MHzOH}
\end{figure*}

\begin{figure*}
\includegraphics[width=15cm]{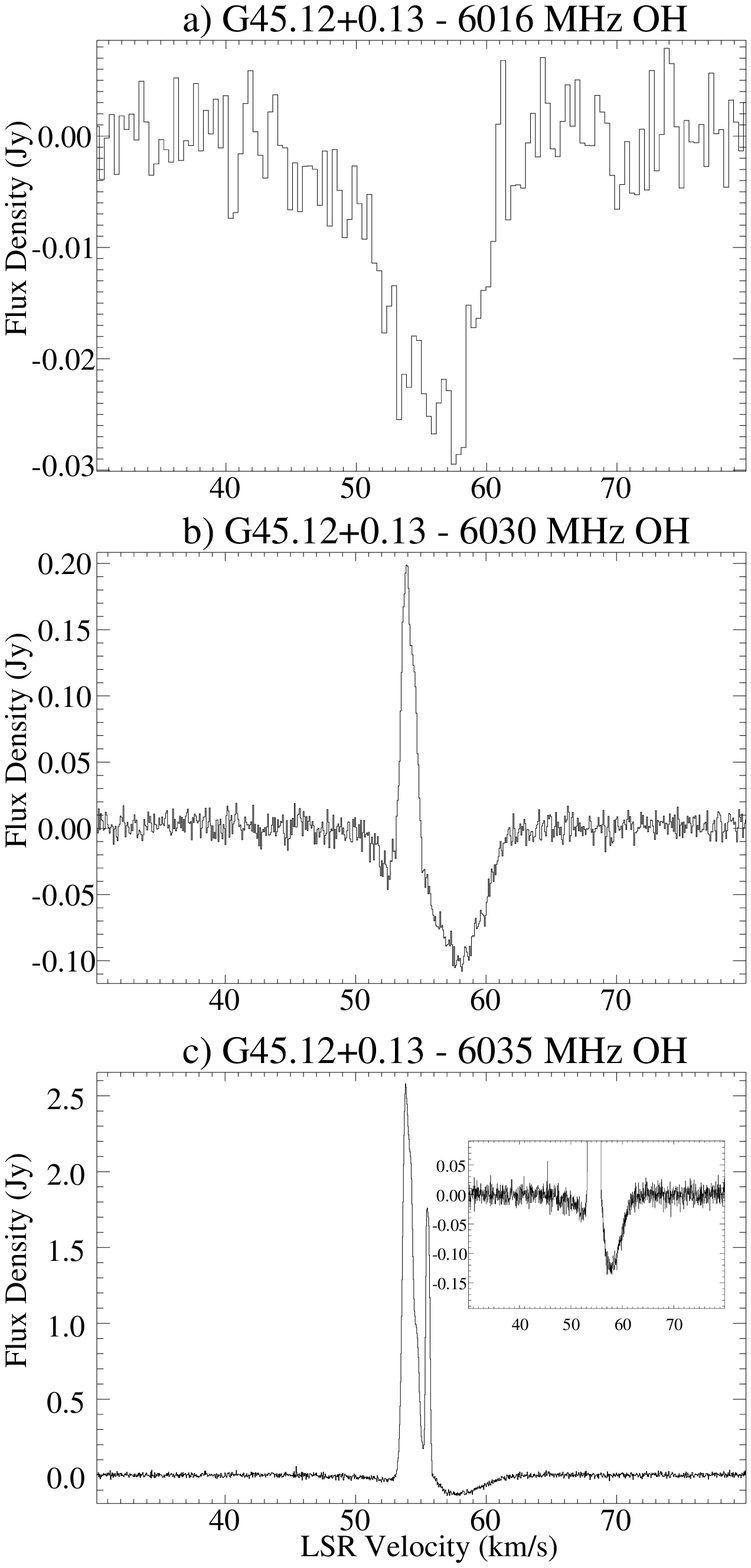}
\caption{Detection of excited OH lines toward G45.12+0.13 at three different frequencies (see Tables~\ref{tb_followup}, \ref{tb_OH6035}, \ref{tb_tentdetec}). The inset in the bottom panel shows a zoom-in view of the absorption line. The data shown in the figure are available as supplemental online material.}
\label{fOHG45}
\end{figure*}

\begin{figure*}
\includegraphics[angle = 0,width=15cm]{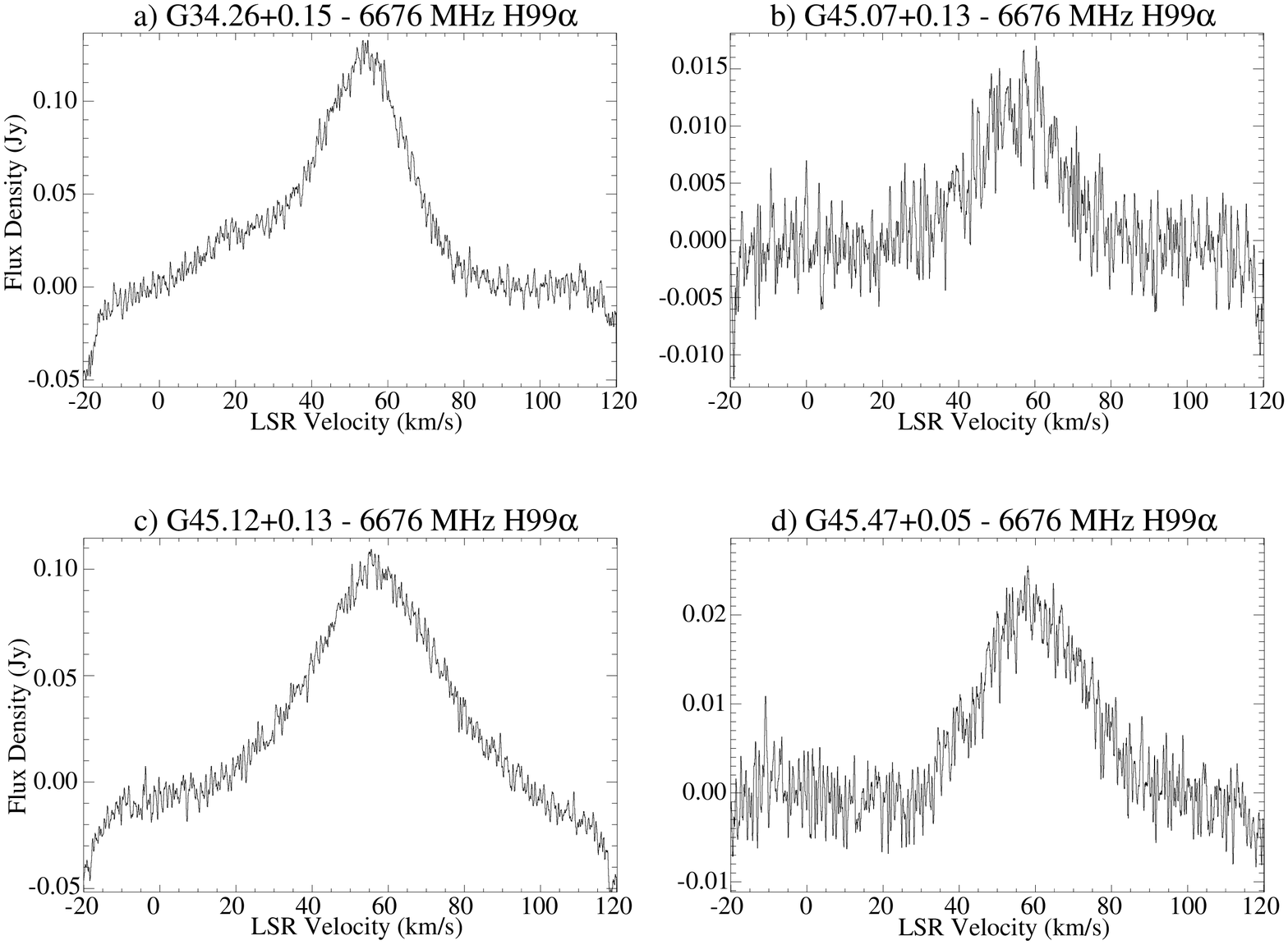}
\caption{Detection of the H99$\alpha$ radio recombination line at 6.676$\,$GHz toward four sources. In the case of G34.26+0.15 and G45.12+0.13, the line parameters could not be reliably measured (see Table~\ref{tb_H99a}). The data shown in the figure are available as supplemental online material.}
\label{RRL}
\end{figure*}

\begin{figure*}
\includegraphics[angle = 90,width=\textwidth]{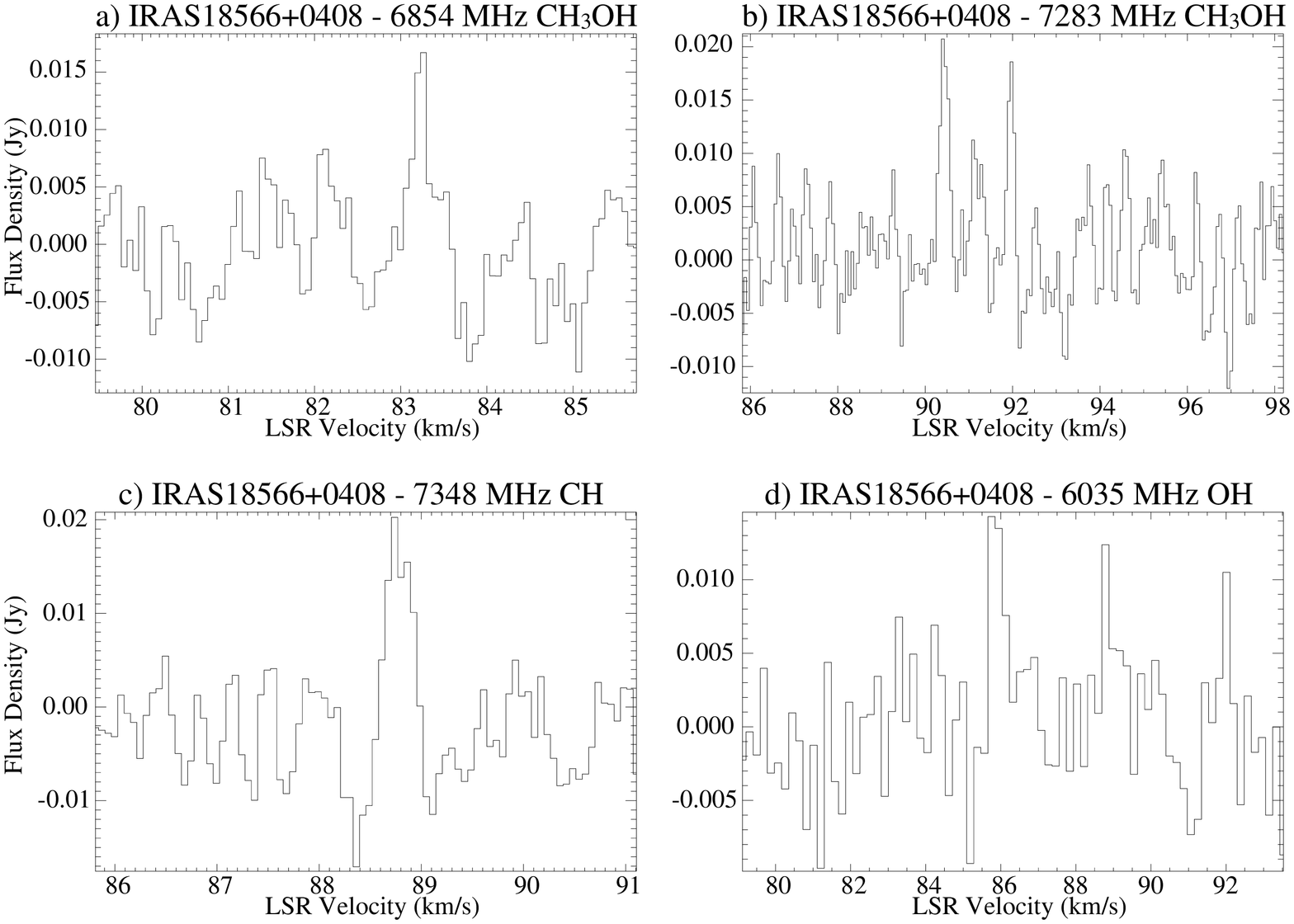}
\caption{Spectra of weak lines toward IRAS$\,$18566+0408. Panels a), b), and c) show tentative detections, while the weak line shown in panel d) is a confirmed 6.035$\,$GHz OH detection, as \citet{Al-Marzouk_2012ApJ...750..170A} reported this line and showed that the line is highly variable, possibly correlated with the variability of 6.7$\,$GHz CH$_3$OH masers in the region. The tentative detections were not confirmed in the follow-up observations (Table~\ref{tb_followup}), thus the lines may be variable or artefacts. The data shown in the figure are available as supplemental online material.}
\label{ftenative_detections}
\end{figure*}

\section{Discussion} \label{Discussion}

\subsection{6.7 GHz CH$_3$OH Absorption}
\label{6.7GHz_Abs}

\subsubsection{Detections Reported in This Work}
\label{6.7GHz_Abs_this_work}

Figures~\ref{fCH3OH} and \ref{f_CH3OH_G45} show our detection of 6.7$\,$GHz methanol lines in nine star-forming regions. Among the sources in the sample, G34.26+0.15 is a well-known region of high-mass star formation that harbours methanol masers as well as absorption. The absorption line is also visible in the spectrum reported by \cite{Breen_2015MNRAS.450.4109B}. The case of G34.26+0.15 exemplifies the challenges in detecting 6.7$\,$GHz CH$_3$OH absorption, as strong masers coincident in velocity could easily mask weak absorption lines (e.g., Figure~\ref{fCH3OH}a) and high-sensitivity observations are needed for detection. For example, \cite{2000A&A...362..723C} studied the 6.7$\,$GHz methanol masers toward the UCH{\small \text{II}} region in G34.26+0.15, but the typical RMS of their observations was $\sim$3.5$\,$Jy, which was too high for detection of the absorption line. G45.47+0.05 is another example where the RMS of previous observations was too high for the detection of weak absorption; \cite{1995MNRAS.272...96C} and \cite{2007ApJ...656..255P} reported masers in this region with a typical RMS of $\sim$60$\,$mJy and $\sim$70$\,$mJy, respectively, which would render undetectable the $-$30$\,$mJy line reported here. 

To explore the physical conditions of the molecular clouds with 6.7$\,$GHz absorption in our sample (Table~\ref{tb_CH3OH-6668}), we focus on G45.12+0.13, as it is the only region whose absorption profile was not affected by overlapping maser lines in at least one epoch (Figure~\ref{f_CH3OH_G45}). G45.12+0.13 harbours a cometary UCH{\small II} region ($\sim 6$\arcsec~mean diameter, \citealt{Wood_1989ApJS...69..831W}) against which NH$_3$ has been detected in absorption \citep{Hofner_1999ApJ...514...899H}. Our data show consistent 6.7$\,$GHz absorption in three different epochs (see Figure ~\ref{f_CH3OH_G45}). We can rule out absorption solely against the CMB (e.g., from molecular material behind or to the side of the UCH{\small II} region but still within the Arecibo beam) based on the size of the Arecibo primary beam and the flux density of the absorption, because there would not be enough CMB flux density to account for the absorption if the cloud were smaller than the Arecibo beam, and an anti-inverted 6.7$\,$GHz CH$_3$OH absorption region more extended than the Arecibo beam (behind the UCH{\small II} region) would be unreasonable based on the physical size of the region and physical conditions needed for anti-inversion \citep{2008AA...484L..43I}. Thus, the absorption is most likely against the radio continuum of the UCH{\small II} region. In the case of local thermodynamic equilibrium (LTE), we obtain a CH$_3$OH column density of $10^{15}\,$cm$^{-2}$ assuming $T_K = 20\,$K (e.g., \citealt{Houghton_1995MNRAS.273.1033H}) and a filling factor of 1\footnote{A filling factor of 1 is assumed given that the absorption has not been imaged.} with respect to the continuum measured with Arecibo (Table~\ref{tb_CH3OH-6668}). Similar column densities are likely for the other 6.7$\,$GHz CH$_3$OH absorption regions in our sample, however significant overlap with maser lines precludes a reliable determination (see Figure~\ref{fCH3OH}). The LTE analysis of 6.7$\,$GHz CH$_3$OH absorption in Sgr B2 \citep{Houghton_1995MNRAS.273.1033H} resulted in a CH$_3$OH column density of 10$^{18}\,$cm$^{-2}$, which is significantly greater than the column density we find for G45.12+0.13. Sgr B2 is one of the most extreme sites of star formation in our Galaxy, thus, a greater column density is reasonable.

\subsubsection{Association with High-Mass Star Forming Regions}
\label{6.7GHz_Abs_HMSFR}

\begin{table*}
\begin{threeparttable} 
\centering
\caption{Literature Review of 6.7$\,$GHz Methanol Absorption} 
\label{tb_literature_review}
\begin{tabular}{lccccccc} 
  \hline
{Reference} 					& {Telescope} 	 & {Beam Size}  & {RMS$^{a}$}	& {Total Number}  & {Abs.$^c$}    & {Emis.}    & {Abs. Detection}\\
{ }         					& {}           	 & {(arcsec)}   & {(mJy)}	& {of Targets$^{b}$}& {}          & {}         & {Rate$^{d}$}   	\\
\hline                                                                                          
This Work		    			& Arecibo        & 43       	& 5		& 12	    	  & 4  		  & 9	       & 33$\%$	\\ 
\cite{Breen_2015MNRAS.450.4109B}$^{e}$      	& Parkes         & 192 	    	& 70		& (265)      	  & 5  		  & 265	       & 2$\%$	\\ 
\cite{1995MNRAS.272...96C}         		& Parkes         & 192      	& 60		& 245             & 11 	          & 245	       & 4$\%$  \\ 
\cite{Caswell_2010MNRAS.404.1029C}$^{e}$	& Parkes         & 192      	& 70		& (183)       	  & 5  		  & 183	       & 3$\%$	\\ 
\cite{2011MNRAS.417.1964C}$^{e}$		& Parkes 	 & 192		& 70		& (198)      	  & 4		  & 198        & 2$\%$	\\ 
\cite{2010AA...517A..56F}      			& Effelsberg 	 & 120		& 50		& 296	  	  & 0		  & 55         & 0$\%$	\\ 
\cite{1993MNRAS.262...43G}      		& Hartebeesthoek & 420      	& 500		& 62	  	  & 2?$^f$	  & 18	       & 3$\%$	\\ 
\cite{2010MNRAS.409..913G}$^{e}$		& Parkes         & 192      	& 70		& (119)      	  & 2  		  & 119	       & 2$\%$	\\ 
\cite{2012MNRAS.420.3108G}$^{e}$		& Parkes         & 192      	& 70		& (207)      	  & 6  		  & 207	       & 3$\%$	\\ 
\cite{2008AA...484L..43I} 			& Effelsberg     & 120      	& 0.6 - 8	& 8     	  & 1  		  & 0	       & 13$\%$	\\ 
\cite{Impellizzeri_2008_PhD}                    & Effelsberg     & 120          & 0.6 - 12      & 10              & 2?$^g$        & 2?$^g$     & 20$\%$ \\ 
\cite{1991ApJ...380L..75M}			& 140ft NRAO     & 300      	& 400		& 123 		  & 8  		  & 80	       & 7$\%$   \\ 
\cite{Olmi_2014AA...566A..18O}			& Arecibo    	 & 43      	& 5     	& 107 		  & 1  		  & 37	       & 1$\%$   \\ 
\cite{2007ApJ...656..255P}        		& Arecibo        & 43       	& 70            & (86)      	  & 1  		  & 86	       & 1$\%$\\ 
\cite{Pandian_2008AA...489.1175P}        	& Arecibo        & 43       	& 1.2		& 5     	  & 2  		  & 0	       & 40$\%$ \\ 
\cite{1997MNRAS.291..261W}           		& Parkes         & 198      	& 100 	        & 535		  & 3  		  & 201	       & 0.6$\%$\\ 
\cite{2008AA...485..729X}			& Effelsberg	 & 120 		& 100           & 89		  & 1 		  & 10	       & 1$\%$\\ 
\cite{2017ApJ...846..160Y}           		& TMRT          & 180      	& 20     	& 1473		  & 0  		  & 12	       & 0$\%$	\\ 
\hline
\multicolumn{8}{p{16.2cm}}{$^{a}\,$ Representative RMS values from the articles are listed. We give a range in cases where RMS values for different sources varied by a factor of $\sim 10$ or more.}\\
\multicolumn{8}{p{16.2cm}}{$^{b}\,$ The column lists the total number of sources observed in a specific survey including non-detections. In case of blind surveys, we list in parenthesis the number of sources with 6.7$\,$GHz CH$_3$OH detection (emission and/or absorption).}\\
\multicolumn{8}{p{16.2cm}}{$^{c}\,$ In many cases, the number of regions with absorption reported in the table should be considered a lower limit, as the presence of absorption was based on inspection of published spectra designed to show masers and not weak absorption features; in addition, not all spectra were shown in the articles (e.g., \citealt{1995PASA...12..186W}, \citealt{1997MNRAS.291..261W}, \citealt{2008AA...485..729X}, \citealt{2017ApJ...846..160Y}).}\\
\multicolumn{8}{p{16.2cm}}{$^{d}\,$ The detection rate in the case of blind surveys is the percentage of sources detected (emission or absorption) that show absorption.}\\
\multicolumn{8}{p{16.2cm}}{$^{e}\,$ Part of the Methanol MultiBeam (MMB) Survey \citep{Green_2009MNRAS.392..783G}. In most cases, spectra shown in MMB papers are from Parkes MX-mode follow-up observations ($\sim 70\,$mJy RMS) instead from the main survey cubes ($\sim 170\,$Jy RMS); e.g., see \cite{Caswell_2010MNRAS.404.1029C}.}\\
\multicolumn{8}{p{16.2cm}}{$^{f}\,$ \citet{1993MNRAS.262...43G} reported absorption toward G338.47$+$0.29, however, the absorption was not detected in MMB data \citep{2011MNRAS.417.1964C}, hence, the absorption feature was likely an artefact.}\\
\multicolumn{8}{p{16.2cm}}{$^{g}\,$ In addition to the detection of 6.7$\,$GHz CH$_3$OH absorption toward NGC$\,$3079, Mrk$\,$348 was listed as a tentative detection. Tentative detection of 6.7$\,$GHz CH$_3$OH emission was reported toward Mrk$\,$3 and NGC$\,$6240. We note that a sub-sample of the observations reported in \cite{Impellizzeri_2008_PhD} were published in \cite{2008AA...484L..43I}, i.e., these are not independent studies.}\\
\end{tabular}
\end{threeparttable} 
\end{table*}

\begin{figure*}
\begin{tabular}{ll}
\includegraphics[width=0.8\textwidth]{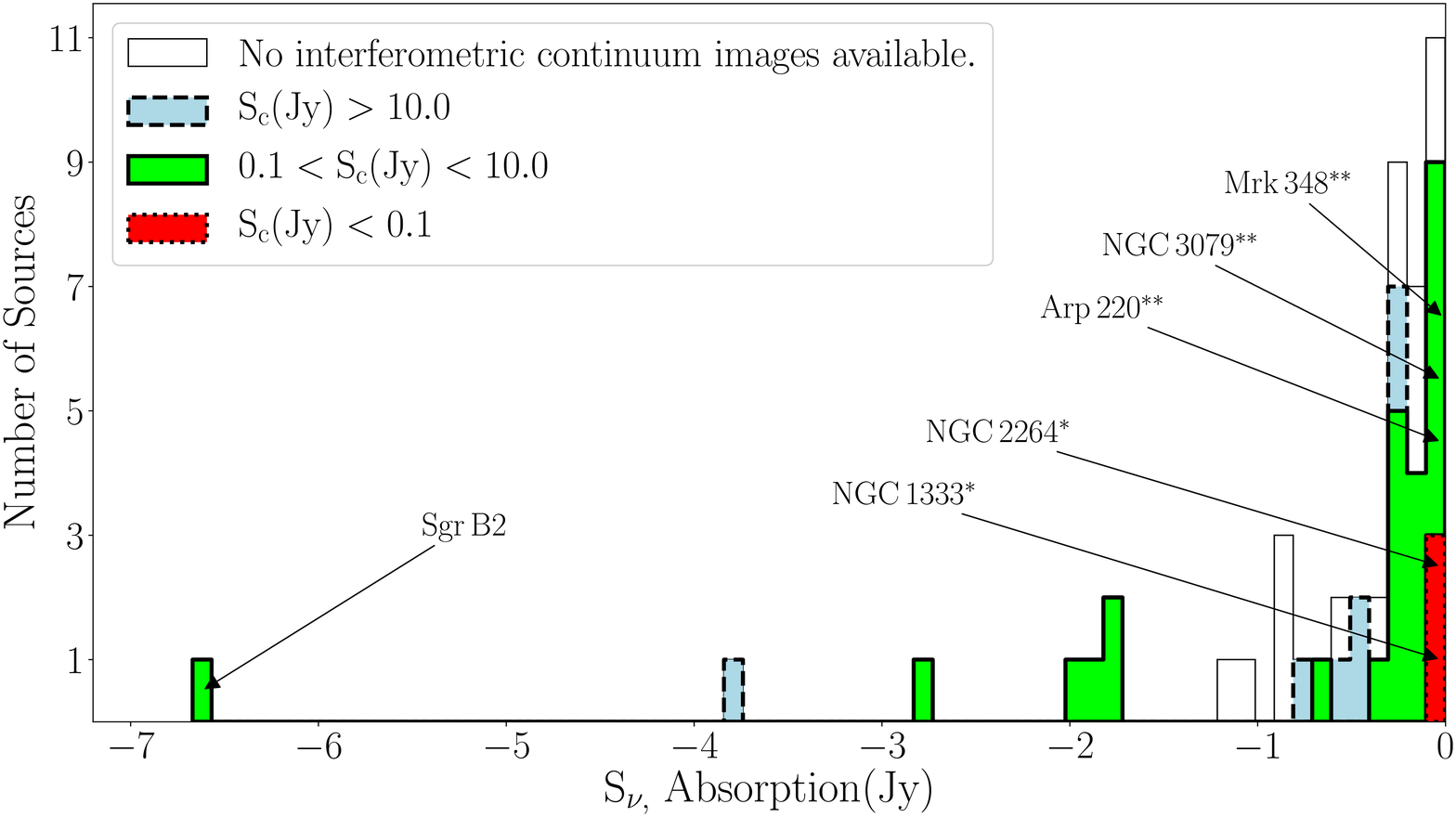} \\
\includegraphics[width=0.8\textwidth]{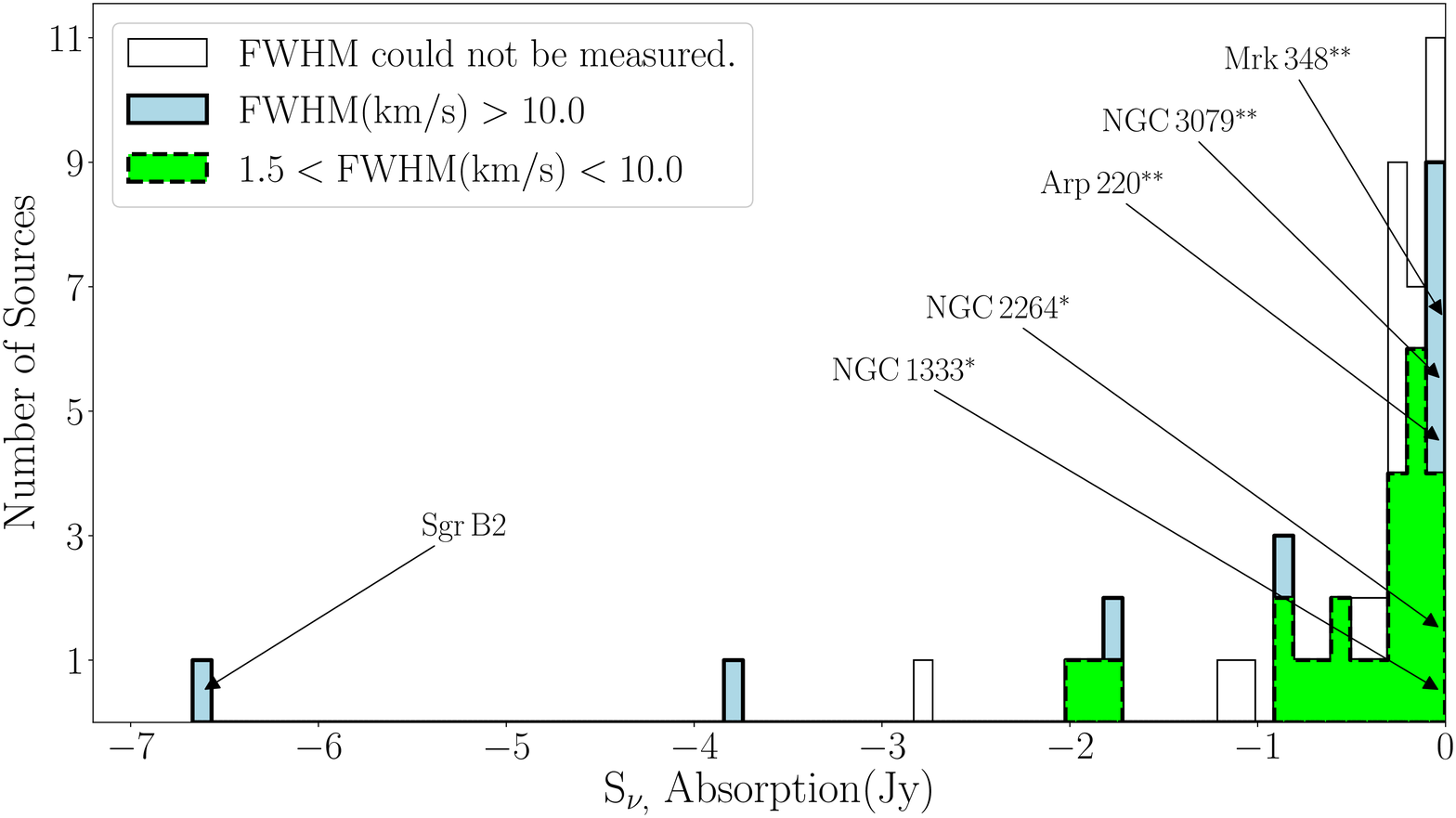}
\end{tabular}
\caption{Histogram of the number of sources with 6.7$\,$GHz methanol absorption at different flux density bins. Colours and line type (solid, dashed, dotted) in the {\it upper panel} identify the radio continuum flux density based on interferometric images. Colours and line type in the {\it lower panel} identify ranges in FWHM of the absorption lines. ($^*$) NGC$\,$1333 is a low-mass star forming region and the absorption is against the CMB \citep{Pandian_2008AA...489.1175P}. The absorption line detected toward NGC$\,$2264 is also likely against the CMB \citep{Pandian_2008AA...489.1175P}. ($^{**}$) Extragalactic absorption has been reported toward NGC$\,$3079, Arp$\,$220 and Mrk$\,$348.}
\label{fFWHM}
\end{figure*}

\begin{figure*}
\begin{tabular}{ll}
\includegraphics[width=0.8\textwidth]{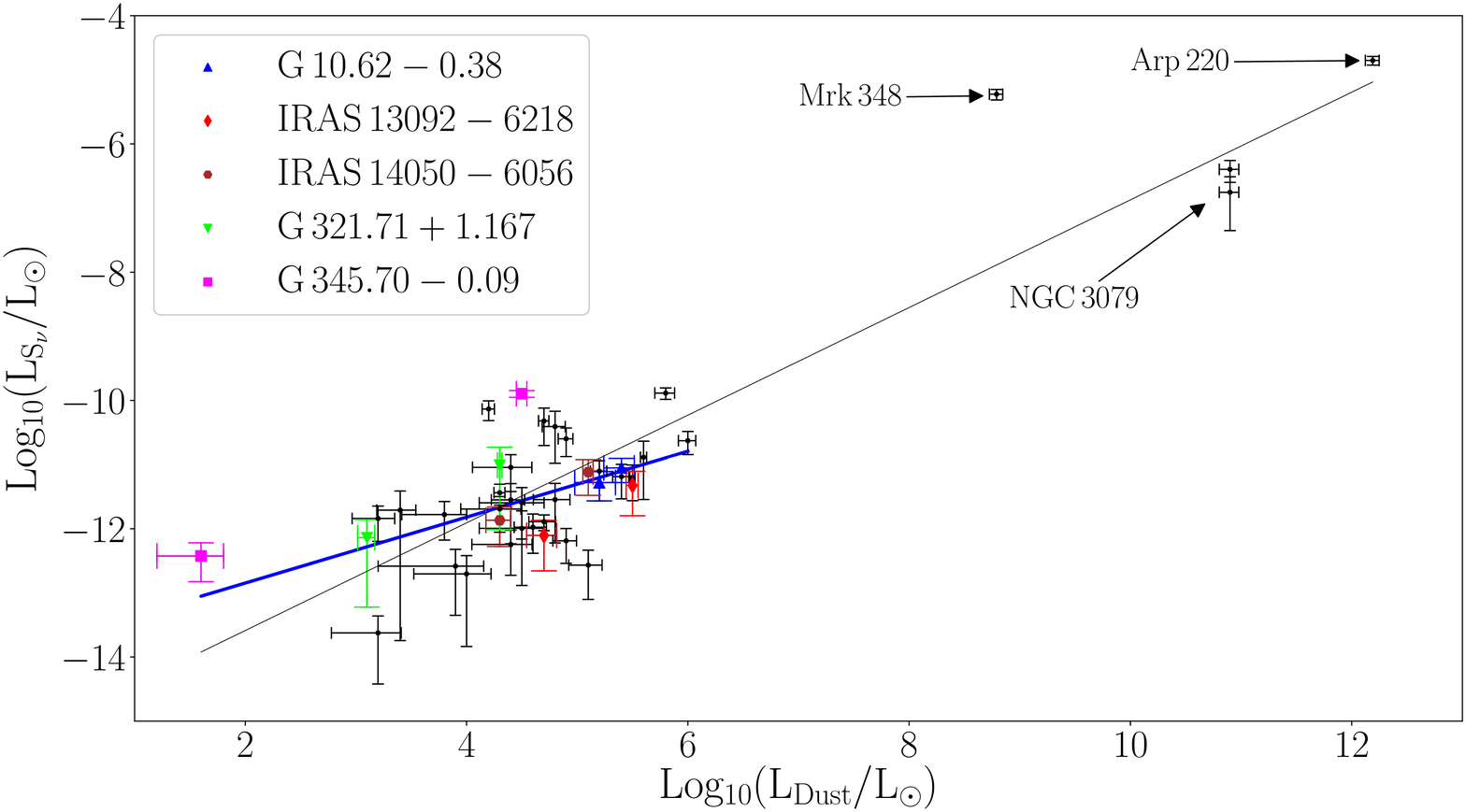} \\
\includegraphics[width=0.8\textwidth]{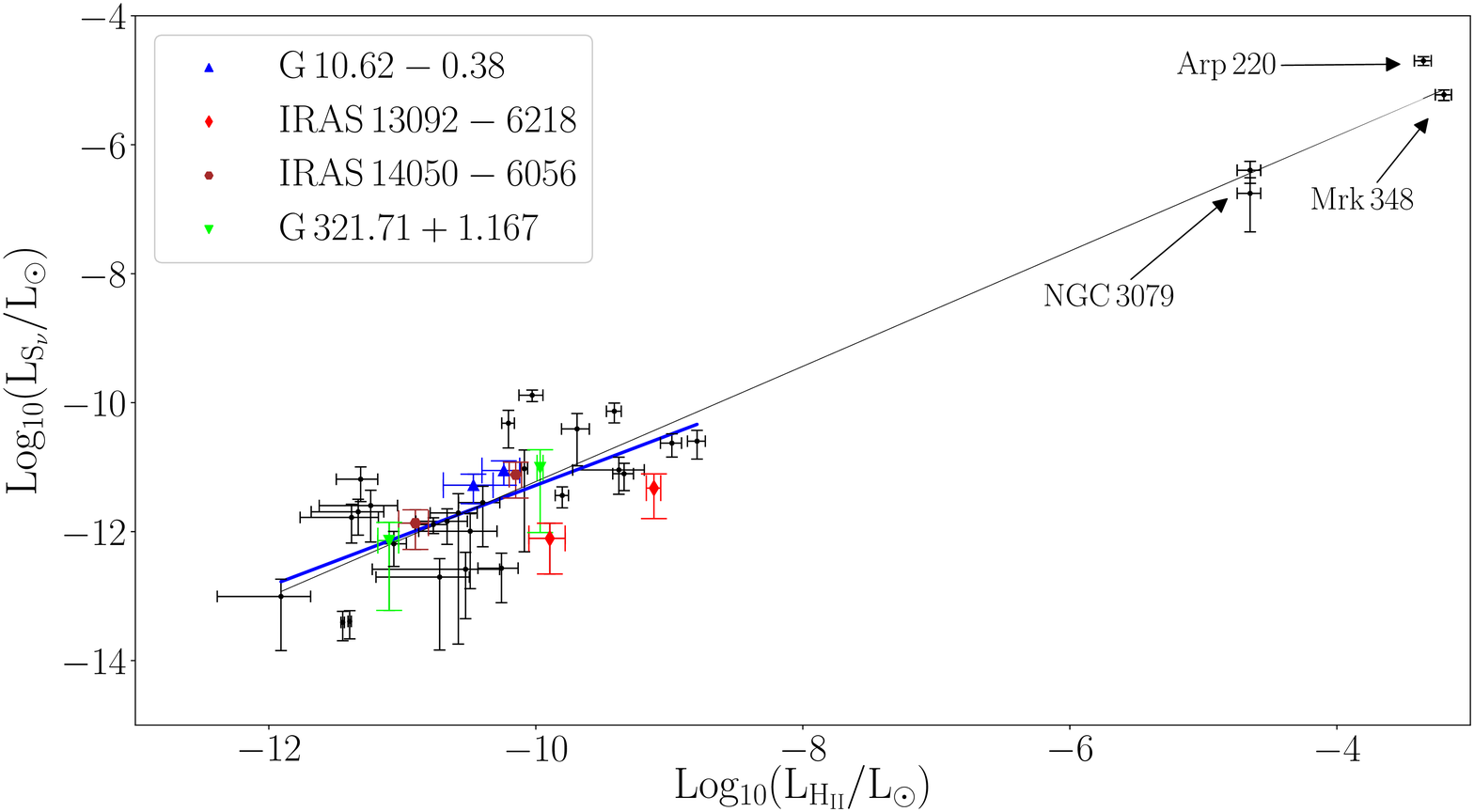}
\end{tabular}
\caption{Absorption luminosity of the 6.7$\,$GHz CH$_3$OH line ($L_{S_\nu} = 4 \pi |S_{\nu, line}| d^2$) as a function of FIR luminosity ({\it upper panel}) and 5$\,$GHz radio luminosity ({\it lower panel}). Sources identified with different symbols and colours are regions for which the distance ambiguity has not been resolved, and thus, significantly different luminosity values are possible. No interferometric continuum observations of G345.70$-$0.09 are available, thus, it was not included in the {\it lower panel}. All data in the {\it upper panel} were fit with an $\alpha = 0.84 \pm 0.07$ power law (black fit); the thick slope (blue line) shows the fit ($\alpha = 0.51 \pm 0.14$) when the extragalactic sources are excluded. The data in the {\it lower panel} were also fit with a power-law ($\alpha = 0.89 \pm 0.05$) when all data are included (black fit), and with a power-law $\alpha = 0.78 \pm 0.16$ when only sources with $log({L_{\text {HII}}/L_{\odot}})$ between $-$12 and $-$8 are included (thick fit line shown in blue).}
\label{luminosity}
\end{figure*}

Most 6.7$\,$GHz CH$_3$OH studies focus on masers, whereas statistical studies of galactic 6.7$\,$GHz CH$_3$OH absorption have not been reported. We conducted an extensive literature review of 6.7$\,$GHz CH$_3$OH surveys to search for absorption features, which may have been discussed in articles, included in tables, or simply shown in published maser spectra without discussion\footnote{An initial literature review of 6.7$\,$GHz CH$_3$OH absorption was included in the MS thesis of \cite{kim2014}.}. Table~\ref{tb_literature_review} shows the results of our literature review. We focus on blind surveys and/or high-sensitivity observations, and not on the large number of low-sensitivity and/or interferometric observations of 6.7$\,$GHz CH$_3$OH masers that were unlikely to detect absorption (e.g., the single-dish surveys by \citealt{Szymczak_2000A&AS..143..269S} and \citealt{Szymczak_2002A&A...392..277S} had 3$\sigma$ detection limits greater than 1$\,$Jy). Also, while there are high-sensitivity blind interferometric surveys for CH$_3$OH masers (e.g., \citealt{2019MNRAS.482.5349R}), high angular resolution observations with extended arrays can resolve out extended absorption, which would not be observable in maser spectra. We also exclude from Table~\ref{tb_literature_review} articles that focus on high-sensitivity studies of single prominent sources, as the detection rates from such studies would not be statistically meaningful. For instance, \cite{Houghton_1995MNRAS.273.1033H} was not included in Table~\ref{tb_literature_review} as they conducted an interferometric study of 6.7$\,$GHz CH$_3$OH emission and absorption of Sgr B2, which had previously been detected in surveys. Likewise, we do not include in the table the extreme high-sensitivity observations with the Arecibo Telescope (0.2$\,$mJy RMS) of Arp$\,$220, which resulted in detection of 6.7$\,$GHz CH$_3$OH absorption \citep{Salter_2008AJ....136..389S}. However, both sources (Sgr B2 and Arp$\,$220) are included in Figures~\ref{fFWHM} and \ref{luminosity}.

The detection rate of methanol absorption in our survey is 33$\%$, which, with the caveat of our relatively small sample, is among the highest detection rates found in the literature (most surveys have detection rates $<10\%$)\footnote{We stress that many of the detection rates listed in Table~\ref{tb_literature_review} were not explicitly reported in the articles but calculated based on spectra shown in the papers, thus, the detection rates should be considered lower limits as it is possible that authors may have detected absorption lines but not shown them in spectra and/or not mentioned them in tables or text.}. We note that the papers with the highest absorption detection rates (\citealt{Pandian_2008AA...489.1175P} and this work) were conducted with the 305m Arecibo Telescope, which is the most sensitive telescope for detection of spectral lines at 6$\,$GHz wavelengths. With the exception of the Effelsberg survey for extragalactic 6.7$\,$GHz CH$_3$OH by \cite{2008AA...484L..43I} (which had an RMS sensitivity between 0.6 and 8$\,$mJy; see also \citealt{Impellizzeri_2008_PhD}), these surveys have among the lowest RMS noise levels, and thus are most suitable to detect weak absorption features.

A graphical summary of the literature review is presented in Figures~\ref{fFWHM} and \ref{luminosity}. Both panels of Figure~\ref{fFWHM} show histograms of number of sources with 6.7$\,$GHz methanol absorption for different bins of flux density absorption. In the upper panel, colours and line styles represent different ranges of compact radio continuum, i.e., flux density values from compact sources (no CMB or extended Galactic emission) in the telescope beam of the CH$_3$OH absorption features. The continuum values are from Table~\ref{tb_CH3OH-6668}, \cite{Purcell_2013ApJS..205....1P}, \cite{Hoare_2012PASP..124..939H}, \cite{1987A&A...171..261C} or interferometric observations from the NRAO VLA Image Archive\footnote{We note that multiple continuum sources could be within the beam of the telescope used for detection of 6.7$\,$GHz CH$_3$OH absorption; e.g., there is a compact, ultracompact and two hypercompact H{\small II} regions in the Arecibo beam of the G34.26+0.15 observations (e.g., \citealt{Sewilo_2004ApJ...605..285S}; \citealt{Gomez_2000RMxAA..36..161G}).}. Most sources where 6.7$\,$GHz absorption lines have been reported harbour bright continuum sources, i.e., Figure~\ref{fFWHM} upper panel is mostly populated by sources highlighted in green (thick solid line) and blue (thick dashed line) colours, i.e., sources with radio continuum greater than 0.1$\,$Jy). Interferometric observations are needed to confirm that the absorption is against the compact radio continuum sources, as previously observed toward Sgr$\,$B2 \citep{Houghton_1995MNRAS.273.1033H}.

The lower panel of Figure~\ref{fFWHM}$\,$ examines the distribution of 6.7$\,$GHz methanol absorption flux density for different full width at half maximum (FWHM) values of the absorption lines. In some cases, the FWHM could not be measured due to overlapping bright masers. FWHM values less than 1.5\kms~could indicate relatively quiescent molecular environments, e.g., pre-stellar cores candidates, while values greater than 1.5$\,$kms$^{-1}$ indicate significant turbulent environments and/or outflows, e.g., regions of active star formation (e.g., \citealt{Bergin_2007ARA&A..45..339B}). No {\it bona fide} detections of narrow absorption lines ($<1.5$\kms) were found in our literature review. For instance, G338.47$+$0.29 has the narrowest methanol absorption-like feature we found in the literature (FWHM $\sim 1$\kms, \citealt{1993MNRAS.262...43G}), however, the absorption was not detected in MMB data \citep{2011MNRAS.417.1964C}, hence, the line was likely an artefact. Thus, the absorption lines in our literature review are broad, and taking into account the likely association with compact radio continuum sources, our study suggests that 6.7$\,$GHz methanol absorption is tracing relatively evolved regions of high-mass star formation, i.e., in the UCH{\small II} region phase.

Figure~\ref{luminosity} shows absorption luminosity density ($L_{S_\nu} = 4 \pi |S_{\nu, line}| d^2$) as a function of FIR luminosity (estimated from Herschel data, \citealt{Molinari_2016A&A...591A.149M}) and radio luminosity density (from interferometric observations). Our compilation of 6.7$\,$GHz CH$_3$OH absorption regions show a weak dependence of absorption flux density with infrared luminosity for Galactic sources (i.e., excluding NGC$\,$3079, Arp$\,$220 and Mrk$\,$348), and a stronger dependence of line absorption with radio continuum luminosity, which again supports the interpretation that the absorption is associated with compact ionized regions instead of absorption against extended Galactic and CMB continuum.

A selection bias is a concern as many surveys (including this one) were targeted observations toward active regions of star formation, hence, association with star formation is expected. Blind maser surveys can easily miss absorption toward weak radio continuum sources (such as low-mass star forming regions) because their sensitivity is appropriate for much stronger features. For instance, \cite{2007ApJ...656..255P} conducted a blind survey with the Arecibo Telescope and absorption was found only toward one region of high-mass star formation; their RMS (70 to 85$\,$mJy) was too high to detect weak absorption features in low-mass star forming regions, such as those reported by \cite{Pandian_2008AA...489.1175P}. Another large scale blind survey of this transition is the 6.7$\,$GHz MMB project (\citealt{2010MNRAS.409..913G}, \citealt{Caswell_2010MNRAS.404.1029C}, \citealt{2011MNRAS.417.1964C}, \citealt{2012MNRAS.420.3108G}, \citealt{Breen_2015MNRAS.450.4109B}), which searched for bright methanol masers across the Galactic plane. The typical RMS of the MMB main survey is $\sim$0.17$\,$Jy and the RMS of the MMB MX follow-ups is $\sim$0.07$\,$Jy, e.g., \cite{Caswell_2010MNRAS.404.1029C}, which precludes detection of weak absorption features. 

The work by \cite{Olmi_2014AA...566A..18O} shows that despite the observational bias, our conclusion that 6.7$\,$GHz CH$_3$OH absorption is tracing mostly regions with compact radio continuum sources, i.e., active sites of high-mass star formation, is reliable. They observed 107 Hi-GAL sources with the Arecibo Telescope with high sensitivity ($\sim 5 - 10\,$mJy RMS) but found only {\it one} possible absorption feature in G59.78+0.63 (see their figure~3). \cite{Olmi_2014AA...566A..18O} does not report radio continuum data toward the observed methanol masers, however, the authors carried out follow-up VLA observations of a smaller sample of sources, that allowed them to map the compact radio emission (Olmi et al. {\it in prep.}). They found that in most cases the peak flux density was $\lesssim 1\,$mJy$\,$beam$^{-1}$ (in one case it was as high as $\sim$26$\,$mJy$\,$beam$^{-1}$). Thus, the combination of high-sensitivity, low radio continuum, and low detection rate supports our conclusion that 6.7$\,$GHz CH$_3$OH absorption mostly traces active star forming regions with significant free-free emission, instead of cold molecular clouds with absorption against the CMB.

\subsection{Excited OH Lines}
\label{ExcitedOH}

Another transition included in our survey is the 6.035$\,$GHz OH line, which has been found in other high-mass star forming regions (e.g., \citealt{2007ApJ...670L.117F}). We detected 6.035$\,$GHz OH masers in five of the twelve regions (one is a very weak detection toward IRAS$\,$18566+0408), and absorption in three of the five sources. To our knowledge, the 6035$\,$MHz OH masers toward G45.07+0.13 are a new detection. Although G45.07+0.13 was observed by \cite{Avison_2016MNRAS.461..136A}, their 3$\sigma$ detection limit of about 0.5$\,$Jy is higher than the $\sim 0.02 - 0.2\,$Jy flux densities that we measured. We list the RMS and the line parameters of the detections in Table~\ref{tb_OH6035} (see also Figures~\ref{fig_6035MHzOH}, \ref{fOHG45} and \ref{ftenative_detections}). Four of the regions have strong radio continuum (Table~\ref{tb_OH6035}), as also confirmed by the detection of hydrogen radio recombination lines (Figure~\ref{RRL}); IRAS$\,$18566+0408 has a weak radio continuum source ($\sim$1$\,$mJy; \citealt{Araya_2005ApJ...618..339A}; \citealt{Hofner_2017ApJ...843...99H}). We discuss in more detail two of the sources here:

\subsubsection{G34.26+0.15}

\cite{Al-Marzouk_2012ApJ...750..170A} reported 6.035$\,$GHz OH masers and a broad 6.035$\,$GHz OH absorption feature in G34.26+0.15.\footnote{{The masers in this source were also detected by \cite{Avison_2016MNRAS.461..136A} at similar flux density levels, but the absorption line was below their sensitivity level.}} The observations reported here confirm this broad absorption. The absorption line is asymmetric, with a prominent low-velocity wing indicative of outflow/expansion motion (see Figure~\ref{fig_6035MHzOH})\footnote{Additional observations of the absorption line were presented in the MS thesis of \cite{kim2014}.}. The G34.26+0.15 region shows strong infall signatures in many common molecular lines (e.g., \citealt{Liu_2013ApJ...776...29L}; \citealt{Wyrowski_2016A&A...585A.149W}) which affect and dilute potential traces of expanding motions. The OH signal we discovered clearly shows a blue-shifted asymmetry indicating expansion and based on VLA observations (article in preparation), the absorption is in front of the cm continuum, which removes any geometric ambiguity when interpreting the line kinematics.

\subsubsection{G45.12+0.13}

In addition to the 6.035$\,$GHz OH emission and absorption lines shown in Figure~\ref{fig_6035MHzOH}, we detected OH lines at 6.016$\,$GHz (absorption) and 6.030$\,$GHz (absorption and emission) toward G45.12+0.13 (Figure~\ref{fOHG45}). The figure also includes a second epoch of the 6.035$\,$GHz OH transition: the spectra from Figure~\ref{fig_6035MHzOH} show data from 2008 while Figure~\ref{fOHG45} shows data from March 2014. It is evident from the figures that the 6.035$\,$GHz absorption profile is consistent between both epochs, but some changes are clear in the maser profile. In particular, the disappearing of the double peak of the brightest maser between 53\kms~and 55\kms~(although the line from March 2014, Figure~\ref{fOHG45}, shows a red asymmetry indicative of overlapping masers). There is also an increase in intensity of the high velocity maser peak at 55.5\kms. Variability of the maser profile is also evident when comparing to the spectrum reported by \cite{Avison_2016MNRAS.461..136A}. 

We detected OH absorption in the three observed 6$\,$GHz OH transitions, with the 6.016$\,$GHz OH line being the weakest. The 6.016$\,$GHz, 6.030$\,$GHz and 6.035$\,$GHz absorption lines are detected at the same velocity, near 58\kms~(Figure~\ref{fOHG45}). As in the case of OH absorption lines detected toward Sgr A East \citep{2007ApJ...670L.117F}, the 6$\,$GHz lines in G45.12+0.13 likely trace thermal absorption against the radio continuum source. Masers were detected at 6.030 and 6.035$\,$GHz at a velocity of $\sim$55\kms, but we did not detect emission of the 6.016$\,$GHz line at a 3$\sigma$ level of 7.5$\,$mJy. Detection of masers at 6.030 and 6.035$\,$GHz, with the 6.035$\,$GHz maser being brighter, and no detection of a 6.016$\,$GHz maser, is expected from theoretical models of OH masers in star forming regions \citep{Cragg_2002MNRAS.331..521C}.

\subsection{Tentative Detections and Non-Detections}
\label{Tentative}

Figure~\ref{ftenative_detections} shows spectra of weak lines tentatively detected toward the high-mass star forming region IRAS$\,$18566+0408 but not confirmed in our follow-up observations. Even though the lines are weak, at least the 6.035$\,$GHz OH transition is a reliable detection, given that it is found at the same velocity as the 6.035$\,$GHz OH maser reported by \cite{Al-Marzouk_2012ApJ...750..170A}. As mentioned in that paper, this 6.035$\,$GHz OH maser is variable, which would explain the low intensity detected in this project. The excited CH$_3$OH lines at 6.854$\,$GHz and 7.283$\,$GHz and the CH line at 7.348$\,$GHz are tentative detections as they have flux densities just above the 3$\sigma$ level. Their LSR velocities compared with detections of other transitions (e.g., \citealt{araya10}) suggest that they are real, although as noted above, these lines were not confirmed in our follow-up observations (Table~\ref{tb_followup}). Given the strong variability of several molecular maser species in IRAS$\,$18566+0408 (\citealt{Al-Marzouk_2012ApJ...750..170A}, \citealt{araya10}), it is possible that the tentative lines are variable masers, although further high-sensitivity observations are needed to confirm this. We note that theoretical work \citep{Sobolev_1997MNRAS.288L..39S} predicts inversion of the 7.283$\,$GHz CH$_3$OH transition. We found no reference to previous detections of these transitions in the literature, e.g., see \cite{Matthews_1986A&A...161..329M} for the case of CH.

As reported in Tables~\ref{tb_config} and \ref{tb_followup}, other transitions of several species like D$_2$CO, CH$_3$CHO, and CH were not detected. These non-detections are not surprising because of abundance constraints, the need of extreme column densities, and/or special physical conditions needed for population inversion. Specifically, non-detections of D$_2$CO are expected as the abundance of double deuterated molecules is low (e.g., \citealt{Turner_1990ApJ...362L..29T}). In the case of CH$_3$CHO, detectable emission of the transition included in this work requires extreme environments (the transition has been only reported toward Sgr$\,$B2, \citealt{Bell_1983A&A...127..420B}; although other CH$_3$CHO transitions have been detected in other sources, see \citealt{Matthews_1985ApJ...290..609M}, \citealt{Ikeda_2001ApJ...560..792I}, \citealt{Chengalur_2003A&A...403L..43C}). Regarding the non-detections of other CH$_3$OH transitions, bright masers of these low-frequency lines require very special conditions for population inversion (e.g., \citealt{Sobolev_1997MNRAS.288L..39S}). In the case of the CH non-detections, the energy levels of the rotational-excited transitions included in this work are unlikely to be significantly populated (see non-detections reported by \citealt{Matthews_1986A&A...161..329M}).

\section{Summary} \label{Summary}

We report one of the highest sensitivity surveys for molecular lines at 6$\,$GHz frequencies conducted to date. The 305$\,$m Arecibo Telescope was used to observe a sample of twelve intermediate and high-mass star forming regions. Molecular transitions of CH, CH$_3$CHO, CH$_3$C$_5$N, CH$_3$OH, D$_2$CO, H$_2$CS, OH, and radio recombination lines were searched for, with channel widths between 0.03 and 0.7\kms~and achieving RMS levels of the order of $\sim 5\,$mJy for most sources and transitions. We confirmed broad absorption of the 6.035$\,$GHz OH transition toward G34.26+0.15. The absorption is asymmetric and shows a blue-shifted velocity wing indicative of expansion (perhaps an outflow) of molecular material. 

Our high-sensitivity observations allowed us to investigate an aspect of the 6.7$\,$GHz CH$_3$OH transition that has been often neglected in the literature, i.e., absorption. We report detection of absorption in four regions of our sample, i.e., a 33$\%$ detection rate, which is among the highest detection rates of 6.7$\,$GHz CH$_3$OH absorption reported in the literature. A literature review of 6.7$\,$GHz CH$_3$OH absorption shows that the known absorption sources are mostly high-mass star forming regions with bright (>0.1$\,$Jy) continuum sources.

\section*{Acknowledgements}

We acknowledge an anonymous referee for comments that greatly improved our manuscript. E.D.A. acknowledges partial support from NSF grant AST-1814063, a Summer Stipend Grant from Western Illinois University, and computational resources donated by the WIU Distinguished Alumnus Frank Rodeffer. P.H. acknowledges partial support from NSF grant AST-1814011. E.D.A. would like to thank WIU AstroLab students for participating in the project, in particular Daniel M. Halbe and Hyung Kwan Kim; also Jos\'e Andr\'es D{\'\i}az Lor{\'\i}a for taking part in the planning stages of one of the telescope proposals. The Arecibo Observatory is operated by the University of Central Florida under a cooperative agreement with the National Science Foundation (AST-1822073), and in alliance with Universidad Ana G. M\'endez and Yang Enterprises. This research has made use of NASA's Astrophysics Data System. The National Radio Astronomy Observatory (NRAO) is a facility of the National Science Foundation operated under cooperative agreement by Associated Universities, Inc.

\section*{Data Availability}

The data underlying this article are available in the article and in its online supplementary material.





\begin{thebibliography}{100}

\bibitem[Al-Marzouk et al.(2012)]{Al-Marzouk_2012ApJ...750..170A} Al-Marzouk, A.~A., Araya, E.~D., Hofner, P., et al.\ 2012, \apj, 750, 170 

\bibitem[Araya et al.(2002)]{Araya_2002ApJS..138...63A} Araya, E., Hofner, P., Churchwell, E., et al.\ 2002, \apjs, 138, 63

\bibitem[Araya et al.(2004)]{Araya_2004ApJS..154..579A} Araya, E., Hofner, P., Linz, H., et al.\ 2004, \apjs, 154, 579 

\bibitem[Araya et al.(2005)]{Araya_2005ApJ...618..339A} Araya, E., Hofner, P., Kurtz, S., et al.\ 2005, \apj, 618, 339 

\bibitem[Araya et al.(2006)]{Araya_2006AJ....132.1851A} Araya, E., Hofner, P., Olmi, L., Kurtz, S., \& Linz, H.\ 2006, \aj, 132, 1851 

\bibitem[Araya et al.(2008)]{Araya_2008ApJ...675..420A} Araya, E., Hofner, P., Kurtz, S., Olmi, L., \& Linz, H.\ 2008, \apj, 675, 420 

\bibitem[Araya et al. (2010)]{araya10} Araya, E. D., Hofner, P., Goss, W. M., Kurtz, S., Richards, A. M. S., Linz, H., Olmi, L., \& Sewi{\l}o, M. 2010, ApJL, 717, 133

\bibitem[Avison et al.(2016)]{Avison_2016MNRAS.461..136A} Avison, A., Quinn, L.~J., Fuller, G.~A., et al.\ 2016, \mnras, 461, 136 

\bibitem[Bell et al.(1983)]{Bell_1983A&A...127..420B} Bell, M.~B., Matthews, H.~E., \& Feldman, P.~A.\ 1983, \aap, 127, 420

\bibitem[Bergin \& Tafalla(2007)]{Bergin_2007ARA&A..45..339B} Bergin, E.~A., \& Tafalla, M.\ 2007, \araa, 45, 339

\bibitem[Beuther et al.(2002)]{Beuther_2002ApJ...566..945B} Beuther, H., Schilke, P., Menten, K.~M., et al.\ 2002, \apj, 566, 945

\bibitem[Breen et al.(2015)]{Breen_2015MNRAS.450.4109B} Breen, S.~L., Fuller, G.~A., Caswell, J.~L., et al.\ 2015, \mnras, 450, 4109 

\bibitem[Caswell \& Haynes(1987)]{1987A&A...171..261C} Caswell, J.~L., \& Haynes, R.~F.\ 1987, \aap, 171, 261 

\bibitem[Caswell et al. (1995)]{1995MNRAS.272...96C} Caswell, J.~L., Vaile, R.~A., Ellingsen, S.~P., Whiteoak, J.~B., \& Norris, R.~P.\ 1995, \mnras, 272, 96 

\bibitem[Caswell et al. (2010)]{Caswell_2010MNRAS.404.1029C} Caswell, J.~L., Fuller, G.~A., Green, J.~A., et al.\ 2010, \mnras, 404, 1029 

\bibitem[Caswell et al.(2011)]{2011MNRAS.417.1964C} Caswell, J.~L., Fuller, G.~A., Green, J.~A., et al.\ 2011, \mnras, 417, 1964  

\bibitem[Cesaroni et al.(2015)]{Cesaroni_2015A&A...579A..71C} Cesaroni, R., Pestalozzi, M., Beltr{\'a}n, M.~T., et al.\ 2015, \aap, 579, A71

\bibitem[Chengalur \& Kanekar(2003)]{Chengalur_2003A&A...403L..43C} Chengalur, J.~N., \& Kanekar, N.\ 2003, \aap, 403, L43

\bibitem[Codella \& Moscadelli(2000)]{2000A&A...362..723C} Codella, C., \& Moscadelli, L.\ 2000, \aap, 362, 723 

\bibitem[Cragg et al.(2002)]{Cragg_2002MNRAS.331..521C} Cragg, D.~M., Sobolev, A.~M., \& Godfrey, P.~D.\ 2002, \mnras, 331, 521 

\bibitem[Ellingsen et al.(1994)]{Ellingsen_1994MNRAS.267..510E} Ellingsen, S.~P., Norris, R.~P., Whiteoak, J.~B., et al.\ 1994, \mnras, 267, 510 

\bibitem[Fish et al.(2007)]{2007ApJ...670L.117F} Fish, V.~L., Sjouwerman, L.~O., \& Pihlstr{\"o}m, Y.~M.\ 2007, \apjl, 670, L117

\bibitem[Fontani et al.(2010)]{2010AA...517A..56F} Fontani, F., Cesaroni, R., \& Furuya, R.~S.\ 2010, \aap, 517, A56 

\bibitem[Gaylard \& MacLeod(1993)]{1993MNRAS.262...43G} Gaylard, M.~J., \& MacLeod, G.~C.\ 1993, \mnras, 262, 43

\bibitem[Goedhart et al.(2004)]{2004MNRAS.355..553G} Goedhart, S., Gaylard, M.~J., \& van der Walt, D.~J.\ 2004, \mnras, 355, 553

\bibitem[G{\'o}mez et al.(2000)]{Gomez_2000RMxAA..36..161G} G{\'o}mez, Y., Rodr{\'\i}guez-Rico, C.~A., Rodr{\'\i}guez, L.~F., et al.\ 2000, \rmxaa, 36, 161

\bibitem[Green et al.(2009)]{Green_2009MNRAS.392..783G} Green, J.~A., Caswell, J.~L., Fuller, G.~A., et al.\ 2009, \mnras, 392, 783 

\bibitem[Green et al.(2010)]{2010MNRAS.409..913G} Green, J.~A., Caswell, J.~L., Fuller, G.~A., et al.\ 2010, \mnras, 409, 913 

\bibitem[Green et al.(2012)]{2012MNRAS.420.3108G} Green, J.~A., Caswell, J.~L., Fuller, G.~A., et al.\ 2012, \mnras, 420, 3108 

\bibitem[Hoare et al.(2012)]{Hoare_2012PASP..124..939H} Hoare, M.~G., Purcell, C.~R., Churchwell, E.~B., et al.\ 2012, \pasp, 124, 939 

\bibitem[Hofner et al. (1999)]{Hofner_1999ApJ...514...899H} Hofner, P., et al.\ 1999, \apj, 514, 899

\bibitem[Hofner et al.(2017)]{Hofner_2017ApJ...843...99H} Hofner, P., Cesaroni, R., Kurtz, S., et al.\ 2017, \apj, 843, 99 

\bibitem[Houghton \& Whiteoak(1995)]{Houghton_1995MNRAS.273.1033H} Houghton, S., \& Whiteoak, J.~B.\ 1995, \mnras, 273, 1033 

\bibitem[Ikeda et al.(2001)]{Ikeda_2001ApJ...560..792I} Ikeda, M., Ohishi, M., Nummelin, A., et al.\ 2001, \apj, 560, 792

\bibitem[Impellizzeri (2008)]{Impellizzeri_2008_PhD} Impellizzeri, C.~M.~V., 2008, PhD thesis, Univ. of Bonn 

\bibitem[Impellizzeri et al. (2008)]{2008AA...484L..43I} Impellizzeri, C.~M.~V., Henkel, C., Roy, A.~L., \& Menten, K.~M.\ 2008, \aap, 484, L43

\bibitem[Kim (2014)]{kim2014} Kim, H. K. 2014, MS Thesis, Western Illinois University

\bibitem[Kurtz et al.(1994)]{Kurtz_1994ApJS...91..659K} Kurtz, S., Churchwell, E., \& Wood, D.~O.~S.\ 1994, \apjs, 91, 659

\bibitem[Kurtz et al.(2000)]{Kurtz_2000prpl.conf..299K} Kurtz, S., Cesaroni, R., Churchwell, E., et al.\ 2000, Protostars and Planets IV, 299

\bibitem[Liu et al.(2013)]{Liu_2013ApJ...776...29L} Liu, T., Wu, Y., \& Zhang, H.\ 2013, \apj, 776, 29

\bibitem[MacLeod et al.(2018)]{MacLeod_2018MNRAS.478.1077M} MacLeod, G.~C., Smits, D.~P., Goedhart, S., et al.\ 2018, \mnras, 478, 1077 

\bibitem[Matthews et al.(1985)]{Matthews_1985ApJ...290..609M} Matthews, H.~E., Friberg, P., \& Irvine, W.~M.\ 1985, \apj, 290, 609

\bibitem[Matthews et al.(1986)]{Matthews_1986A&A...161..329M} Matthews, H.~E., Bell, M.~B., Sears, T.~J., et al.\ 1986, \aap, 161, 329

\bibitem[Menten(1991)]{1991ApJ...380L..75M} Menten, K.~M.\ 1991, \apjl, 380, L75

\bibitem[Miralles et al.(1994)]{Miralles_1994ApJS...92..173M} Miralles, M.~P., Rodriguez, L.~F., \& Scalise, E.\ 1994, \apjs, 92, 173

\bibitem[Molinari et al.(2016)]{Molinari_2016A&A...591A.149M} Molinari, S., Schisano, E., Elia, D., et al.\ 2016, \aap, 591, A149

\bibitem[Olmi et al.(2014)]{Olmi_2014AA...566A..18O} Olmi, L., Araya, E.~D., Hofner, P., et al.\ 2014, \aap, 566, A18 
 
\bibitem[Pandian et al.(2007)]{2007ApJ...656..255P} Pandian, J.~D., Goldsmith, P.~F., \& Deshpande, A.~A.\ 2007, \apj, 656, 255 

\bibitem[Pandian et al.(2008)]{Pandian_2008AA...489.1175P} Pandian, J.~D., Leurini, S., Menten, K.~M., Belloche, A., \& Goldsmith, P.~F.\ 2008, \aap, 489, 1175

\bibitem[Pandian et al.(2011)]{2011ApJ...730...55P} Pandian, J.~D., Momjian, E., Xu, Y., Menten, K.~M., \& Goldsmith, P.~F.\ 2011, \apj, 730, 55

\bibitem[Peng \& Whiteoak(1991)]{Peng_1991PASAu...9..287P} Peng, R., \& Whiteoak, J.~B.\ 1991, Proceedings of the Astronomical Society of Australia, 9, 287  

\bibitem[Purcell et al.(2013)]{Purcell_2013ApJS..205....1P} Purcell, C.~R., Hoare, M.~G., Cotton, W.~D., et al.\ 2013, \apjs, 205, 1   

\bibitem[Rajabi et al.(2019)]{Rajabi_2019MNRAS.484.1590R} Rajabi, F., Houde, M., Bartkiewicz, A., et al.\ 2019, \mnras, 484, 1590

\bibitem[Rickert et al.(2019)]{2019MNRAS.482.5349R} Rickert, M., Yusef-Zadeh, F., \& Ott, J.\ 2019, \mnras, 482, 5349

\bibitem[Rosero et al.(2016)]{Rosero_2016ApJS..227...25R} Rosero, V., Hofner, P., Claussen, M., et al.\ 2016, \apjs, 227, 25

\bibitem[Salter et al.(2008)]{Salter_2008AJ....136..389S} Salter, C.~J., Ghosh, T., Catinella, B., et al.\ 2008, \aj, 136, 389 

\bibitem[Sewilo et al.(2004)]{Sewilo_2004ApJ...605..285S} Sewilo, M., Churchwell, E., Kurtz, S., et al.\ 2004, \apj, 605, 285

\bibitem[Sjouwerman et al.(2010)]{Sjouwerman_2010ApJ...724L.158S} Sjouwerman, L.~O., Murray, C.~E., Pihlstr{\"o}m, Y.~M., Fish, V.~L., \& Araya, E.~D.\ 2010, \apjl, 724, L158 

\bibitem[Sobolev et al.(1997)]{Sobolev_1997MNRAS.288L..39S} Sobolev, A.~M., Cragg, D.~M., \& Godfrey, P.~D.\ 1997, \mnras, 288, L39

\bibitem[Sridharan et al.(2002)]{Sridharan_2002ApJ...566..931S} Sridharan, T.~K., Beuther, H., Schilke, P., et al.\ 2002, \apj, 566, 931

\bibitem[Strack et al.(2019)]{Strack_2019ApJ...878...90S} Strack, A., Araya, E.~D., Lebr{\'o}n, M.~E., et al.\ 2019, \apj, 878, 90

\bibitem[Szymczak et al.(2000)]{Szymczak_2000A&AS..143..269S} Szymczak, M., Hrynek, G., \& Kus, A.~J.\ 2000, \aaps, 143, 269

\bibitem[Szymczak et al.(2002)]{Szymczak_2002A&A...392..277S} Szymczak, M., Kus, A.~J., Hrynek, G., et al.\ 2002, \aap, 392, 277

\bibitem[Szymczak et al.(2018)]{2018MNRAS.474..219S} Szymczak, M., Olech, M., Sarniak, R., Wolak, P., \& Bartkiewicz, A.\ 2018, \mnras, 474, 219  

\bibitem[Tan (2017)]{Tan2017} Tan, W. S. 2017, MS Thesis, Western Illinois University

\bibitem[Turner(1990)]{Turner_1990ApJ...362L..29T} Turner, B.~E.\ 1990, \apjl, 362, L29

\bibitem[Walmsley et al.(1988)]{Walmsley_1988AA...197..271W} Walmsley, C.~M., Batrla, W., Matthews, H.~E., \& Menten, K.~M.\ 1988, \aap, 197, 271 

\bibitem[Walsh et al.(1995)]{1995PASA...12..186W} Walsh, A.~J., Lyland, A.~R., Robinson, G., Bourke, T.~L., \& James, S.~D.\ 1995, \pasa, 12, 186 

\bibitem[Walsh et al.(1997)]{1997MNRAS.291..261W} Walsh, A.~J., Hyland, A.~R., Robinson, G., \& Burton, M.~G.\ 1997, \mnras, 291, 261 

\bibitem[Watson et al.(2003)]{Watson_2003ApJ...587..714W} Watson, C., Araya, E., Sewilo, M., et al.\ 2003, \apj, 587, 714

\bibitem[Williams et al.(2004)]{Williams_2004A&A...417..115W} Williams, S.~J., Fuller, G.~A., \& Sridharan, T.~K.\ 2004, \aap, 417, 115

\bibitem[Wood \& Churchwell(1989)]{Wood_1989ApJS...69..831W} Wood, D.~O.~S., \& Churchwell, E.\ 1989, \apjs, 69, 831 

\bibitem[Wyrowski et al.(2016)]{Wyrowski_2016A&A...585A.149W} Wyrowski, F., G{\"u}sten, R., Menten, K.~M., et al.\ 2016, \aap, 585, A149

\bibitem[Xu et al.(2008)]{2008AA...485..729X} Xu, Y., Li, J.~J., Hachisuka, K., et al.\ 2008, \aap, 485, 729

\bibitem[Xu et al.(2009)]{Xu_2009A&A...507.1117X} Xu, Y., Voronkov, M.~A., Pandian, J.~D., et al.\ 2009, \aap, 507, 1117

\bibitem[Yang et al.(2017)]{2017ApJ...846..160Y} Yang, K., Chen, X., Shen, Z.-Q., et al.\ 2017, \apj, 846, 160 

\end{thebibliography}








\bsp	
\label{lastpage}
\end{document}